\documentclass[11pt]{article}
\usepackage{float}
\usepackage{authblk}
\usepackage{amssymb,amsmath}
\usepackage{graphicx}
\usepackage{caption}
\usepackage{subcaption}
\usepackage{cite}
\usepackage[normalem]{ulem}
\usepackage{color}
\usepackage{setspace}
\usepackage[colorlinks=true
,urlcolor=blue
,citecolor=blue
,linkcolor=blue
,pagecolor=blue
,linktocpage=true
,pdfproducer=medialab
]{hyperref}
\usepackage[a4paper,width=17cm,height=25cm]{geometry}
\usepackage{morefloats}
\usepackage{siunitx}
\usepackage{multirow}

\makeatletter \renewcommand{\@dotsep}{10000} \makeatother
\title{Chirality structure of vector like new physics operators in charged current transitions}
\author[a] {Sajawal Zafar$^*$}
\author[a,b] {Qazi Maaz Us Salam\footnote{These authors contributed equally to this work.}
	\thanks{\texttt qazi.salam@lums.edu.pk (corresponding author)}}
\author[a]{Rana Khan}
\author[a]{Ishtiaq Ahmed}
\author[b] {Rizwan Khalid}
	
\affil[a]{National Center for Physics,  Shahdra Valley Road, Islamabad 44000, Pakistan}
\affil[b]{School of Science and Engineering, Lahore University of Management Sciences (LUMS), Opposite Sector U, D.H.A, Lahore 54792, Pakistan}
\date{}
\begin{document}
\maketitle
\begin{abstract}
We investigate the cascade decay 
$B^{*0}_{s} \rightarrow 
   D_s^-(\rightarrow \tau^-\,\bar\nu_{\tau})\,
\ell^{+}\,{\nu}_\ell$
induced by flavor changing charged currents in the context of the Standard Model and in vector-like couplings beyond the Standard Model. We employ the helicity amplitude formalism for analysis and highlight the role of new vector-like couplings in charged current interactions. We find, in particular, that while new left handed chiral-vector like interactions contribute to the branching ratio, they do not affect the forward-backward asymmetry, or the angular observables. On the other hand, the right handed chiral vector-like coupling in the case of this decay contributes to the branching ratio, forward-backward asymmetry and the angular observables. We confirm that this difference in behavior between the left and right handed NP couplings is a general feature of charged current processes with a vector meson going to a pseudoscalar at the tree level in effective weak theory by cross checking with the cascade decays $B^{*+}_c \rightarrow P(\rightarrow P'\,\mu^+\,\nu_{\mu}) 
\,
\ell^{+}\,{\nu}_\ell$ where $P$ is $B_s^0$ ($D^0$) and $P'$ is $D_s^{*-}$ ($K^-$).
\end{abstract}

%=======================================================================
%                     Introduction
%=======================================================================

\section{Introduction}
A central objective in contemporary particle physics is the investigation of potential signatures indicative of physics beyond the Standard Model (SM), which we refer to as new physics (NP) in 
this manuscript. The SM has demonstrated remarkable success in accurately describing the fundamental particles and their interactions, and 
despite comprehensive searches conducted at major experimental facilities such as the LHC \cite{LHCb:2012myk, CMS:2017zyp}, Belle \cite{Belle-II:2010dht}, and BaBar \cite{Cuhadar-Donszelmann:2001dwm}, no evidence for additional particles that must 
accompany NP scenarios has been discovered.
Several experimental observations however do show anomalies, \emph{i.e.} deviations from the theoretical expectations calculated within the SM framework~\cite{Belle:2024cis,LHCb:2017lpt,LHCb:2012de,LHCb:2018rym}.   
If these anomalies are confirmed as being statistically significant, they shall serve as compelling evidence of physics beyond the SM.
Most of these anomalies belong to the B meson sector.

The flavor changing neutral and charged current semileptonic decays of $B$ mesons provide an ideal testing ground for NP scenarios. In this context, the flavor changing neutral current (FCNC) semileptonic decays of heavy mesons (see, for example, $B \rightarrow K^{(*)} \mu^+ \mu^-$ \cite{LHCb:2013zuf, LHCb:2014cxe, LHCb:2016ykl, LHCb:2015dla, Matias:2012xw, Descotes-Genon:2013vna, LHCb:2020lmf}, $B_0 \rightarrow \phi \mu^+ \mu^-$ \cite{Horgan:2013pva, CDF:2011grz, LHCb:2021zwz}) have emerged as particularly promising probes. Owing to their sensitivity to the underlying neutral current dynamics, these decays offer a fertile testing ground for NP contributions \cite{Vardani:2024bae, Zhang:2024hkn, DAlise:2024qmp, Marshall:2023aje}. 
Extensive efforts have been undertaken to explore possible manifestations of NP, with numerous theoretical scenarios proposed to account for deviations in the flavor-changing neutral current transitions $b \to s$ and $b \to d$ \cite{Biswas:2020uaq, Biswas:2022lhu, Ray:2022eat,Salam:2024nfv,Ball:2004rg,Barman:2018jhz,Aarfi:2025qcp}.  

Recent measurements of semileptonic B-meson decays have shown significant $2\sigma–4\sigma$ discrepancies, particularly in the flavor changing charged current (FCCC) transition \( b \to c \ell \bar{\nu} \). These anomalies are, particularly, highlighted in the observables such as the \( R(D^{(*)}) \), \( R(J/\psi) \) and tau polarization asymmetries~\cite{Blanke:2018yud,Fedele:2022iib,Blanke:2019qrx,Yasmeen:2024cki,Dutta:2018jxz,Li:2016vvp,Sakaki:2013bfa,FermilabLattice:2014ysv,Ligeti:2016npd,Belle:2019ewo,Belle:2019gij,LHCb:2017vlu,Kamali:2018fhr,Colangelo:2016ymy}. Similarly, precise determinations of the CKM matrix element \( |V_{cs}| \) through global analyses of exclusive decays involving the \( c \to s \ell^+ \nu \) transition~\cite{Bolognani:2024cmr,Fajfer:2015ixa,Leng:2020fei,Zhang:2025tki,BESIII:2023gbn,Liu:2021qio,Colangelo:2021dnv,Belfatto:2019swo} are crucial for testing CKM unitarity and probing NP scenarios. These studies collectively offer complementary avenues to scrutinize the SM and explore possible NP contributions.

The decays of vector B mesons ( $B^*_{u,d,s,c}$)~\cite{Wang:2024cyi,Chang:2019xtj,Ray:2019gkv,Chang:2016gyw} remain relatively unexplored owing to their low production rates and detection efficiencies, thereby, presenting significant experimental challenges compared to their pseudoscalar counterparts ($ B_{u,d,s,c}$). However, the increasing data yields from experiments such as LHCb and Belle II offer promising prospects for the future investigation of these vector meson decays~\cite{Wang:2024cyi}.
It is important to mention here that vector mesons predominantly decay through strong and electromagnetic interactions, leading to very short lifetimes and correspondingly suppressed weak-decay branching fractions. For instance, the branching fraction of  
$B_s^{*}$ via weak processes is of the order of $10^{-7}\,\mathrm{GeV}$~\cite{Wang:2024cyi}, which implies that any weak transition such as
$
B_s^{*0} \to D_s^{-}(\to \tau^{-}\bar{\nu}_{\tau})\,\ell^{+}\nu_{\ell}
$
will yield a relatively small signal compared to the dominant strong channel 
$B_s^{*0} \to B_s^{0}\gamma.$
We provide here the theoretical predictions for these rare weak decays in order to identify observables that could become accessible once sufficient statistics are collected.
As far as the experimental feasibility is concerned, one would require facilities capable of producing and reconstructing large samples of vector-$B$ mesons with excellent vertex and lepton-identification capabilities. One of the most promising venues is the LHCb Upgrade~II, which will operate with high-luminosity $pp$ collisions at the HL-LHC and is expected to collect substantial numbers of excited $B_s^{(*)}$ and $B_c^{*}$ states, see the LHCb Upgrade~II~\cite{LHCb:2018roe,LHCb:2020pet}. Additionally, Belle~II with a clean $e^{+}e^{-}$ environment and precise vertexing can provide an alternative approach for studying semileptonic $B^{*}$ decays, albeit with smaller production yields~\cite{Belle-II:2018jsg}.
Consequently, their study may provide an independent and complementary avenue, not only for probing the dynamics of heavy-flavor weak decays, but also for investigating the nature and potential structure of physics beyond the SM. In this context, studies have been conducted on the radiative decays of vector \( B \) mesons~\cite{Wang:2024cyi, Mandal:2020htr, Sakaki:2013bfa,Chang:2015jla, Chang:2016cdi}. Additionally, some theoretical studies on the weak decays of vector \( B \) mesons within the SM framework~\cite{Chang:2016cdi, Grinstein:2015aua, Wang:2012hu, Chang:2016eto, Chang:2015ead} and in the presence of NP scenarios~\cite{Chang:2018sud, Mahata:2024kxu, Mahata:2023geq, Wei:2023rzs, Karmakar:2023rdt, Sheng:2022peg, Sheng:2021iss, Zhang:2019hth,Huang:2025kof} have also been performed. 

{The hadronic form factors for the $B^*_s \to D_s$, $B^*_c \to B_s$, and $B_c^* \to D$ transitions used in our analysis are computed using the covariant light-front quark model (CLFQM) which offers unique advantages compared with other quark-model approaches~\cite{Wang:2024cyi}. The light-front wave functions do not depend on the hadron’s momentum and are therefore Lorentz invariant by construction in this framework. By virtue of this construction, CLFQM is able to treat the final state meson at maximum recoil $q^2=0$ in a relativistic fashion wherein the non-relativistic quark model may not work well. We note that CLFQM has been successfully extended to analyze the transition form factors and hadronic weak decays~\cite{Zhang:2023ypl,Sun:2023uyn,Sun:2023iis,Yang:2024qij}.}
In the pursuit of testing the SM and probing potential signals of NP, it is crucial to focus on observables that are minimally affected by hadronic uncertainties. Among such observables, angular distributions stand out as particularly clean and theoretically robust, offering a sensitive probe for both validating the SM and identifying deviations that may indicate NP effects. In the present study, we, investigate angular observables associated with the FCCC cascade decay: $B^{*0}_{s} \rightarrow 
   D_s^-(\rightarrow \tau^-\,\bar\nu_{\tau})\ell^{+}\,{\nu}_\ell$ and 
   $B^{*+}_{c} \rightarrow 
 P^0(\rightarrow P'\,\mu^+\,\nu_{\mu})
\ell^{+}\,{\nu}_\ell$. The analysis aims to exploit the angular observables of this decay channel as a probe for potential NP effects. We also investigate the forward-backward asymmetry ($A_{FB}$) and the branching ratio (Br). We have done this computation in the SM and in NP scenarios 
involving weak effective chiral left and right handed vector like couplings beyond the SM in a weak effective theory framework.  

Our analysis shows that the $A_{FB}$ and the angular observables are affected only by the right handed vector couplings while the values of the Br vary for both left and right handed vector type couplings. In order to confirm this as a trend for FCCC decays from vector mesons to pseudoscalars, we confirm by looking at the $c\to s$ transition in the decays  $B^{*+}_{c} \rightarrow 
 P^0(\rightarrow P'\,\mu^+\,\nu_{\mu})
\ell^{+}\,{\nu}_\ell$ where $P$ is $B_s^0$ ($D^0$) and $P'$ is $D_s^{*-}$ ($K^-$).

We now comment on the structure of the rest of this manuscript. In Section~\ref{section II} we outline the theoretical formalism that we have used to study the various physical observables in the decay $B^{*0}_{s} \rightarrow D_s^-(\rightarrow \tau^-\,\bar\nu_{\tau})\ell^{+}\,{\nu}_\ell$. We have presented the parameterization of the hadronic matrix elements in terms of relevant form factors and employed the helicity formalism to derive the Br and the four-fold differential decay width. Additionally, we have given the formulae of the $A_{FB}$ and various angular observables, namely $A_{I_{c1}}$, $A_3$, $A_4$, $A_5$, and $A_{6s}$. In Section~\ref{section III}, we begin by giving the numerical inputs and describing the various NP scenarios that we consider. We then go on to give the results of our computation in terms of semi-analytical expressions and discuss the physical observables. We present our results for the observables as a function of the momentum transfer $q^2$ in addition to giving the results integrated over various $q^2$ bins. We also give bar plots and correlation plots in order to facilitate distinguishing characteristics of the various considered NP scenarios. We note, in particular that the left handed NP coupling does not contribute to the $A_{FB}$ or the angular observables in the case of this decay. In order to confirm that this trend extends to other FCCC processes when considered at the tree level in weak effective theory, we then briefly comment on the decays $B^{*+}_{c} \rightarrow 
 P^0(\rightarrow P'\,\mu^+\,\nu_{\mu})\ell^{+}\,{\nu}_\ell$ where $P$ is $B_s^0$ ($D^0$) and $P'$ is $D_s^{*-}$ ($K^-$). Finally, in Section~\ref{section V}, we have summarized our results and the overall conclusions of this study.

\section{Formalism}\label{section II}
The quark level transition that we focus on is $b\to c$ for which the most general effective Hamiltonian contains scalar, vector and tensor type interactions. However, in the present analysis, we restrict our consideration to analyzing only the vector-like 
interactions. This results in the effective Hamiltonian:
\begin{equation}
\mathcal{H}_{\text{eff}} = \frac{4 G_F}{\sqrt{2}} V_{cb} \left[ (1 + C_{V_L}) \mathcal{O}_{V_L}+ C_{V_R} \mathcal{O}_{V_R}\right],
\label{hamiltonian}
\end{equation}
where $G_F$ is the Fermi coupling constant and the four-fermion operators $\mathcal{O}_{V_{L,R}}$ are given by, 
\begin{equation}
   \mathcal{O}_{V_L} = (\bar{c}_L \gamma^\mu b_L)(\bar{\tau}_L \gamma_\mu \nu_{L}) , \qquad
\mathcal{O}_{V_R} = (\bar{c}_R \gamma^\mu b_R)(\bar{\tau}_L \gamma_\mu \nu_{L}). 
\end{equation}
\begin{figure}[H]
    \centering
    \includegraphics[width=0.35\linewidth]{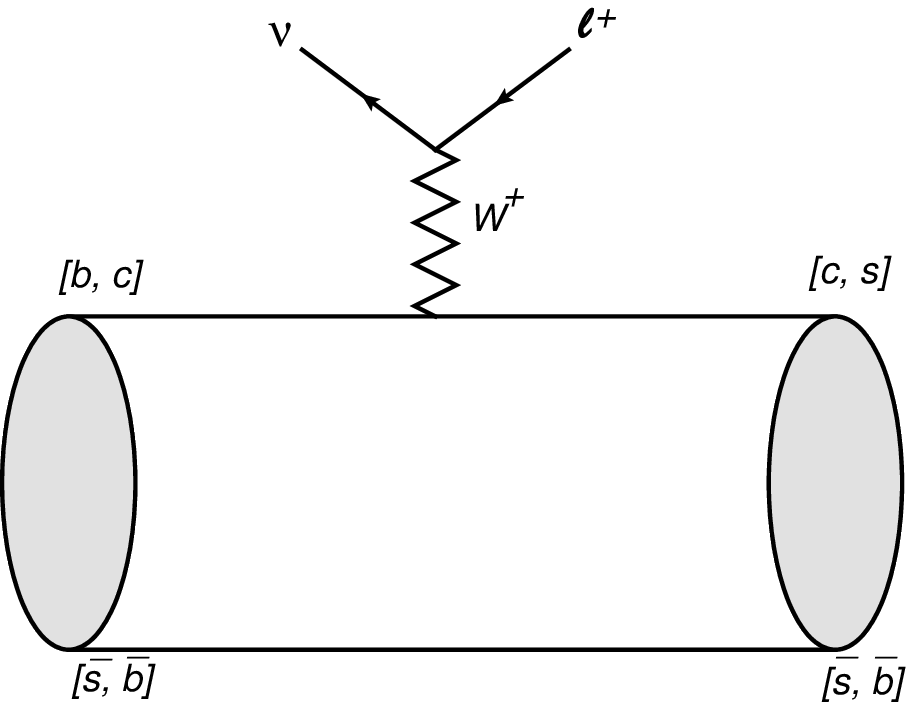}
    \caption{Feynman diagram for semi leptonic $B^{*0}_{s} \rightarrow 
   D_s^-(\rightarrow \tau^-\,\bar\nu_{\tau})\ell^{+}\,{\nu}_\ell$ and 
   $B^{*+}_{c} \rightarrow 
 P^0(\rightarrow P'\,\mu^+\,\nu_{\mu})
\ell^{+}\,{\nu}_\ell$ transitions.}
    \label{fig:enter-label}
\end{figure}

The Wilson coefficients $C_{V_{L,R}}$ parametrize possible NP contributions to this decay. It is to be noted that 
while we focus on the meson decay $B^{*0}_{s} \rightarrow 
   D_s^-(\rightarrow \tau^-\,\bar\nu_{\tau})\ell^{+}\,{\nu}_\ell$, this effective Hamiltonian can be used to study a variety of 
related charged current processes. 

\subsection{Hadronic matrix element}
The hadronic matrix element for the transition of $B^{*0}_{s} \rightarrow D^-_{s}$ is defined via Bauer-Stech-Wirble (BSW) form factors \cite{ Bauer:1986bm, Wang:2024cyi}:

\begin{align}
&\langle D_{s}^-(p'') | \left(V_\mu - A_\mu\right) | B^{*0}_{s}(p', \epsilon) \rangle = 
-\epsilon_{\mu\nu\alpha\beta} \epsilon_{B^{*0}_{s}}^\nu q^\alpha p^\beta 
\frac{V(q^2)}{m_{B^{*}_{s}} + m_{D_{s}}} \nonumber \\
& -i \frac{2 m_{B^{*}_{s}}\epsilon_{B^{*0}_{s}} \cdot q}{q^2} q_\mu A_0(q^2) 
- i \epsilon_{\mu B^{*0}_{s}} (m_{B^{*}_{s}} + m_{D_{s}}) A_1(q^2) - i \frac{\epsilon_{B^{*0}_{s}} \cdot q}{m_{B^{*}_{s}} + m_{D_{s}}} p_\mu A_2(q^2) 
\nonumber\\
&+ i \frac{2 m_{B^{*}_{s}} \epsilon_{B^{*0}_{s}} \cdot q}{q^2} q_\mu A_3(q^2),
\label{eq:decay}
\end{align}
where $p = p'+ p'', q = p'- p''$, and we use the convention $\epsilon_{0123} = 1$.  Here $p'(p'')$ denotes the four-momentum of the initial (final) meson, while $V_{\mu}$ and $A_{\mu}$ are the usual vector and axial-vector currents, respectively. Finally, the scalar functions $V(q^2)$, $A_{0}(q^2)$, $A_{1}(q^2)$ and $A_{2}(q^2)$ are hadronic form factors which are the primary source of hadronic uncertainties.

{The hadronic form factors for the transitions used in this analysis are computed using CLFQM calculations as reported in Ref.~\cite{Wang:2024cyi,Cheng:2003sm,jaus1990semileptonic,jaus1999covariant}. 
All computations are done within the $q^+=0$ reference frame with the form factors available only at spacelike momentum transfers $q^2 = -q^2_\perp \leq 0$. One must obtain the form factors in the timelike region relevant for the physical decay processes. Accordingly, we use the following double-pole approximation to parametrize the form factors obtained in the spacelike region and then extended to the timelike region~\cite{Wang:2024cyi}.
\begin{equation}\label{ff eq}
F(q^2) = \frac{F(0)}{1 - a\left(\frac{q}{m}\right)^2 + b\left(\frac{q}{m}\right)^4}
\end{equation}
where $F$ represents the form factors $V(q^2)$, $A_0(q^2)$, $A_1(q^2)$, and $A_2(q^2)$ and $m$ is the initial meson mass. The values of $F(0)$, and the parameters $a$ and $b$ are obtained by performing a three-parameter fit to the form factors in the range $-(m_{B^*}-m_P)^2\,\mathrm{GeV}^2 \le q^2 \le 0$, with the subscript $P$ representing the final state which in our case are the $B_s$ and $D_{(s)}$.

The form factor values that we use are explicitly provided in Table~2 and Table~13 along with the uncertainties arising from the decay constants of the initial and final mesons. In order to validate our results for stability, we have cross checked our predictions against branching ratio calculations using different form factor parametrizations available in the literature, \emph{i.e.} the Bethe-Salpeter (BS) method~\cite{Wang:2018ryc} and Bauer-Stech-Wirbel (BSW) model~\cite{Chang:2016cdi}. Our calculation yields that there are quantitative differences between various form factor parametrizations  (which is expected and reflected in the systematic uncertainties), but the order of magnitude of all predictions remains consistent ($10^{-7}$ for muon channel, $10^{-8}$ for tau channel). More importantly, the relative ordering and hierarchy of decay rates are preserved across different form factor schemes. This demonstrates that our predictions are stable under variations in the form factor parametrization, and the qualitative physics conclusions are reliable. All our results (Figures in the section~3) explicitly show the impact of form factor uncertainties as bands around the central values. The width of these bands demonstrate how form factor uncertainties affect our predictions for branching ratios and other observables across the full $q^2$ range. One can see the forward-backward asymmetry and angular observables are not much affected by the uncertainty, as is expected. }

\subsection{Two-fold differential decay width}
The two-fold decay amplitude can be written in terms of helicity amplitudes with the decay width given by:
\begin{align}
&\frac{d\Gamma(B^{*0}_{s} \rightarrow  D_s^-
\,
\ell^+\,{\nu}_\ell)}{dq^2}\nonumber\\ &= 
\frac{G_F^2 |V_{cb}|^2 q^2}{192 \pi^3 m_{B^{*}_{s}}^3} \sqrt{\lambda(q^2)} 
\left( 1 - \frac{m_\ell^2}{q^2} \right)^2 \times \Bigg\{ \nonumber \\
&\quad \left(|1 + C_{V_L}|^2 + |C_{V_R}|^2 \right) 
\left[ 
\left( 1 - \frac{m_\ell^2}{q^2} \right) \left(H_{V,+}^2 + H_{V,-}^2 + H_{V,0}^2 \right)
+ \frac{3m_\ell^2}{2q^2} H_{V,t}^2
\right] \nonumber \\
&\quad - 2 \text{Re}\left[(1 + C_{V_L}) (C_{V_R})^*\right]
\left[
\left( 1 + \frac{m_\ell^2}{2q^2} \right) H_{V,0} (H_{V,+} + H_{V,-})
- \frac{3m_\ell^2}{2q^2} H_{V,t}^2
\right] \Bigg\}.
\label{decay rate Bc} 
\end{align}

The helicity amplitudes can be written in terms of form factors as:
\begin{align}
H_{V,\pm}(q^2)
&=(m_{B^{*}_{s}} +m_{D_s}) A_1(q^2) \mp \frac{\sqrt{\lambda(q^2)}}{(m_{B^{*}_{s}} + m_{D_s})} V(q^2), \\
H_{V,0}(q^2)
&= \frac{(m_{B^{*}_{s}} +m_{D_s})}{2  m_{D_s} \sqrt{q^2}} 
\left[
-(m_{B^{*}_{s}}^2 - m_{D_s}^2 - q^2) A_1(q^2) 
+ \frac{\lambda(q^2)}{(m_{B^{*}_{s}} +m_{D_s})^2} A_2(q^2)
\right], \\
H_{V,t}(q^2)
&= -\sqrt{\frac{\lambda(q^2)}{q^2}} A_0(q^2).
\label{hardonic matrix}
\end{align}
where $\lambda(q^2) = \left[(m_{B^{*}_{s}}-m_{D_s})^2-q^2\right]\left[(m_{B^{*}_{s}} +m_{D_s})^2-q^2\right]$.

\subsection{Four-fold Differential Decay width}

The four-fold differential decay distribution for the $B^{*0}_{s} \rightarrow 
\bigl[
   D_s^-(\rightarrow \tau^-\,\bar\nu_{\tau})
\bigr]\,
\ell^{+}\,{\nu}_\ell$ decays can be written as follows \cite{Mandal:2020htr}:
\begin{align}
&\frac{d^4\Gamma}{dq^2\, d\cos\theta_l\, d\cos\theta_{({ D_{s}^-})}\, d\phi} 
\equiv I(q^2, \theta_{l}, \theta_{{ D_{s}^-}}, \phi) \nonumber \\
&= \frac{9}{32\pi}\Big\{
  \left(I_{1s} \sin^2\theta_{{ D_{s}^-}} + I_{1c} \cos^2\theta_{{ D_{s}^-}}\right)+ \left(I_{2s} \sin^2\theta_{{ D_{s}^-}}+ I_{2c} \cos^2\theta_{{ D_{s}^-}}\right)\cos 2\theta_\ell
 \nonumber\\
 &+ \left(I_3 \cos 2\phi + I_9 \sin 2\phi\right)\sin^2\theta_{{ D_{s}^-}}\sin^2\theta_\ell + \left(I_4 \cos\phi + I_8 \sin\phi\right)\sin 2\theta_{{ D_{s}^-}}\sin 2\theta_\ell \nonumber\\
 &+ \left(I_5 \cos\phi + I_7 \sin\phi\right)\sin\theta_\ell + \left(I_{6s} \sin^2\theta_{{ D_{s}^-}} + I_{6c} \cos^2\theta_{{ D_{s}^-}}\right)\cos\theta_\ell
\Big\}.
\end{align}
The angular coefficients $I_{i}’s$ are functions of $q^2$ that incorporate contributions from both short and long-distance physics, where $i=1,2,\dots8$.  
{ The angles appearing in the four-fold differential decay distribution are defined as $\theta_\ell$ is the angle between the momentum of the charged lepton $\ell$ and the direction opposite to the $D_s^{-}$ meson, measured in the dilepton rest frame. $\theta_{D_s^-}$ is the helicity angle of the secondary decay $D_s^{-}\to \tau^{-}\,\bar\nu_{\tau}$. It is defined as the angle between the momentum of the daughter lepton ($\tau^{-}$) and the direction opposite to the parent $B_s^{*0}$ meson, measured in the $D_s^{-}$ rest frame. $\phi$ is the azimuthal angle between the decay planes of the hadronic system ($D_s^{-}\to \tau^{-}\,\bar\nu_{\tau}$) and the dilepton system ($\ell^{+}\nu_{\ell}$), measured in the $B_s^{*0}$ rest frame. These angles $(\theta_\ell,\theta_{D_s^-},\phi)$ completely describe the kinematics of the cascade decay $B_s^{*0}\;\to\; D_s^{-}\bigl(\to \tau^{-}\,\bar\nu_{\tau}\bigr)\,\ell^{+}\nu_{\ell}\,.$
}
The explicit expressions for the angular coefficients in terms of transversity amplitudes $A^L_0$, $A_{\perp,k}$, and $A_t$ are as follows:
\begin{align}
I_{c1} &= N_F \Bigg[ 2 \left( 1 + \frac{m^2_\ell}{q^2} \right) \Big(|A_0^L|^2 \Big) 
 + \frac{4 m_{\ell}^2}{q^2}|A_{tP}^L|^2\Bigg], \\
I_{1s} &= N_F \Bigg[\frac{1}{2} \left( 3 + \frac{m^2_\ell}{q^2} \right) \Big(|A_\perp^L|^2 + |A_\parallel^L|^2\Big) 
\\
I_{2c} &= -2 N_F \left( 1 - \frac{m^2_\ell}{q^2} \right) \Big(|A_0^L|^2\Big),
\\
I_{2s} &= \frac{1}{2} N_F \left( 1 - \frac{m^2_\ell}{q^2} \right) \Big(|A_\perp^L|^2 + |A_\parallel^L|^2\Big)\\
I_3 &= N_F \left( 1 - \frac{m^2_\ell}{q^2} \right) \Big(|A_\perp^L|^2 + |A_\parallel^L|^2\Big),\\
I_4 &= \sqrt{2} N_F \left( 1 - \frac{m^2_\ell}{q^2} \right) \operatorname{Re}[A_0^L (A_\parallel^L)^*], \\
I_5 &= 2\sqrt{2} N_F \Bigg[\operatorname{Re} \Big[(A_0^L]- \frac{m_\ell^2}{q^2}\operatorname{Re}\Big[ (A_{t,P}^{L})^*\big(A_{\parallel}^{L}\big)\Big]\Bigg], \\
I_{6c} &= N_F \frac{8 m_{\ell}^2}{q^2} \operatorname{Re}\big[A_{t,P}^L(A_0^L)\big], \\
I_{6s} &= 4 N_F \operatorname{Re}[(A_\parallel^L)(A_\perp^L)^*],
\end{align}
where,
\begin{equation*}
N_F = \frac{G_F^2 |V_{cb}|^2}{2^{7}3\pi^{3} m_{B^{*}_{s}}^3} q^2 \lambda^{1/2}(q^2) 
\left( 1 - \frac{m_{\ell}^2}{q^2} \right)^2 \mathcal{B}(D_{s}^-(\rightarrow \tau\nu_{\tau})).
\end{equation*}
Where the value of $\mathcal{B}(D_{s}^-(\rightarrow \tau\nu_{\tau})$ is $5.3\%$.
The transversity amplitudes can be expressed in terms of NP WCs as:
\begin{align}
A_0^L &= H_{V,0} \left( 1 + C_{V_L} - C_{V_R} \right), &
A_\parallel^L &= \frac{1}{\sqrt{2}} \left( H_{V,+} + H_{V,-} \right) \left( 1 + C_{V_L} - C_{V_R} \right), \nonumber\\
A_\perp^L &= \frac{1}{\sqrt{2}} \left( H_{V,+} - H_{V,-} \right) \left( 1 + C_{V_L} + C_{V_R} \right), &
A_t^L &= H_{V,t} \left( 1 + C_{V_L} - C_{V_R} \right),\nonumber \\
A_{t,P}^{L} &= A_t^{L}. &
&
\end{align}
It is convenient to focus on the differential decay rate after integrating over $\theta_l$, $\theta_{D_s^-}$ and $\phi$: 
\begin{equation}
    \frac{d\Gamma}{dq^2}=\frac{1}{4}\left(3I_{c1}+6I_{1s}-I_{2c}-2I_{s2} \right). 
\end{equation}
The branching ratio for the decay then becomes:
\begin{equation}
    \text{Br} = \frac{\int dq^2 {d\Gamma}/{dq^2}}{\Gamma_{\text{tot}{({B^{*0}_s}})}},
\end{equation}
where $\Gamma_{\text{tot}(B^{*0}_s)} \cong 0.094\times 10^{-6}~\text{GeV}$ is the total decay width for $B_s^{*0}$~\cite{Wang:2024cyi}.  

In addition to the decay rate, we have calculated the forward-backward asymmetry, $A_{FB}$, and the angular observables, $A_i$ ($i=3,4,5,6s$), which are related to the angular coefficients via the following relations\footnote{We have also looked at the longitudinal helicity fraction, $f_L$, and various lepton polarization asymmetries but have found that these are not sensitive to the vector type NP effects that we address here. Therefore, we have chosen to ignore these observables.}:
\begin{align}
A_{FB} &= \frac{3}{8} \frac{I_{6c} + 2 I_{6s}}{\Gamma_f},&A_3 &= \frac{I_3}{\Gamma_f}, & A_4 &= \frac{2 I_4}{ \pi\Gamma_f}, &
A_5 &= \frac{3}{4} \frac{I_5}{\Gamma_f}, & A_{6s} &= \frac{-27}{8} \frac{I_{6s}}{\Gamma_f}, & A_{I_{c1}} &= \frac{I_{c1}}{\Gamma_f},
\end{align}
where $\Gamma_f\equiv\frac{d\Gamma}{dq^2}$.

\section{Phenomenology}\label{section III}
We now move on to discuss the phenomenology of the $B^{*0}_s \to D^{-}_s (\to \tau^{-}\nu_\tau) l^+ \nu_l$ decay in terms of 
the physical observables of branching fraction, forward-backward asymmetry and angular observables as described in Section~\ref{section II}. The numerical inputs for the analysis are given in Section~\ref{numerical}. 
\subsection{Numerical Inputs}\label{numerical}
The values of parameters which are used in the numerical calculation of the observables are listed in Table \ref{tab:masses}. 
\begin{table}[H]
\centering
\begin{tabular}{c@{\,=\,}c@{\hspace{1cm}} c@{\,=\,}c @{\hspace{1cm}} c@{\,=\,}c @{\hspace{1cm}} c@{\,=\,}c}
\hline
\( m_b \) & 4.8 & \( m_c \) & 1.4 & \( m_s \) & 0.37 & \( m_{e} \) & 0.000511 \\
\( m_{\mu} \) & 0.106 & \( m_\tau \) & 1.776 & \( m_B \) & 5.27965 & \( m_{D} \) & 1.86966 \\
\( m_{D_s} \) & 1.96835 & \( m_{B_s} \) & 5.36688 & \( m_{B^*} \) & 5.32470 & \( m_{B^*_s} \) & 5.4154 \\
\hline
\hline
\end{tabular}
\caption{Numerical values of the input parameters in GeV\(^{-2}\) \cite{Wang:2024cyi}.}
\label{tab:masses}
\end{table}
Table \ref{ff table 1} presents the form factors for the  $B^*_s \to D_s$ transition, evaluated at $q^2 = 0$, along with the parameters $a$ and $b$, which appear in the expression of the extrapolation of the form factors given in Eq.~\ref{ff eq} with the initial mass $m=m_{B^*_s }$. The table includes the vector form factor $V(q^2)$ and axial form factors $A_0(q^2)$, $A_1(q^2)$ and $A_2(q^2)$.

\begin{table}[H]
\centering
{
\begin{tabular}{|c|c|c|c|}
\hline
\textbf{$F(q^2)$} & \textbf{$F(0)$}  & \textbf{$a$} & \textbf{$b$}   \\
\hline
$V(q^2)$   & $0.76^{+0.01+0.01}_{-0.01-0.01}$  & $0.74^{+0.13+0.10}_{-0.13-0.10}$ & $1.62^{+0.02+0.03}_{-0.03-0.03}$\\\hline
$A_0(q^2)$   & $0.63^{+0.00+0.01}_{-0.00-0.01}$  & $0.47^{+0.13+0.11}_{-0.13-0.11}$ & $0.72^{+0.02+0.02}_{-0.03-0.02}$ \\\hline
$A_1(q^2)$   & $0.66^{+0.00+0.01}_{-0.00-0.01}$  & $0.39^{+0.12+0.12}_{-0.12-0.12}$ & $0.56^{+0.01+0.02}_{-0.02-0.02}$\\\hline
$A_2(q^2)$    & $0.56^{+0.00+0.01}_{-0.00-0.00}$   & $0.66^{+0.11+0.11}_{-0.10-0.11}$ & $1.36^{+0.01+0.02}_{-0.01-0.02}$\\ \hline\hline
\end{tabular}}
\caption{Form factors for $B^*_s \to D_s$ transition with uncertainties \cite{Wang:2024cyi}. The first uncertainty represents statistical errors, while the second corresponds to systematic uncertainties.}
\label{ff table 1}
\end{table}

To explore the sensitivity of NP, we have used the numerical values of new WC's which are tabulated in Table \ref{tab:combined}.

\begin{table}[H]
\centering
\begin{minipage}[t]{0.45\linewidth}
\centering
\begin{tabular}{|l|c|c|}
\hline
 & $ C_{V_L}$ & $ C_{V_R}$ \\
\hline
EFT ($>$ 10 TeV) & 0.32 (0.09) & 0.33 (0.09) \\
LQ (4 TeV)       & 0.36 (0.10) & 0.40 (0.10) \\
LQ (2 TeV)       & 0.42 (0.12) & 0.51 (0.15) \\
\hline\hline
\end{tabular}
\end{minipage}
\caption{The ranges for $C_{V_L}$ and $C_{V_R}$ 
in the effective field theory (EFT) and leptoquark model (LQ)
for the $b \to c$ transition~\cite{Iguro:2024hyk}.}
\label{tab:combined}
\end{table}

In this regard, we have considered the following benchmark points: 
\begin{table}[H]
    \centering
    \resizebox{\textwidth}{!}{
    \begin{tabular}{|c|c|c|c|c|c|}\hline
         BI& BII&BIII&BIV&BV&BVI \\\hline
    $C_{V_L}=0.32\, (0.09)$&$C_{V_R}=0.33 \,(0.09)$& $C_{V_L}=0.36 \,(0.10)$& $C_{V_R}=0.40\, (0.10)$&$C_{V_L}=0.42\, (0.12)$&$C_{V_R}=0.51\, (0.15)$\\\hline\hline
    \end{tabular}}
     \caption{The benchmark points for $C_{V_L}$ and $C_{V_R}$ 
in the EFT and LQ model for the $b \to c$ transition.}
    \label{benchmark}
\end{table}

\subsection{Exploration of NP in $b\to c$ transition}

\begin{figure}[H]
\centering
\includegraphics[width=0.3\textwidth]{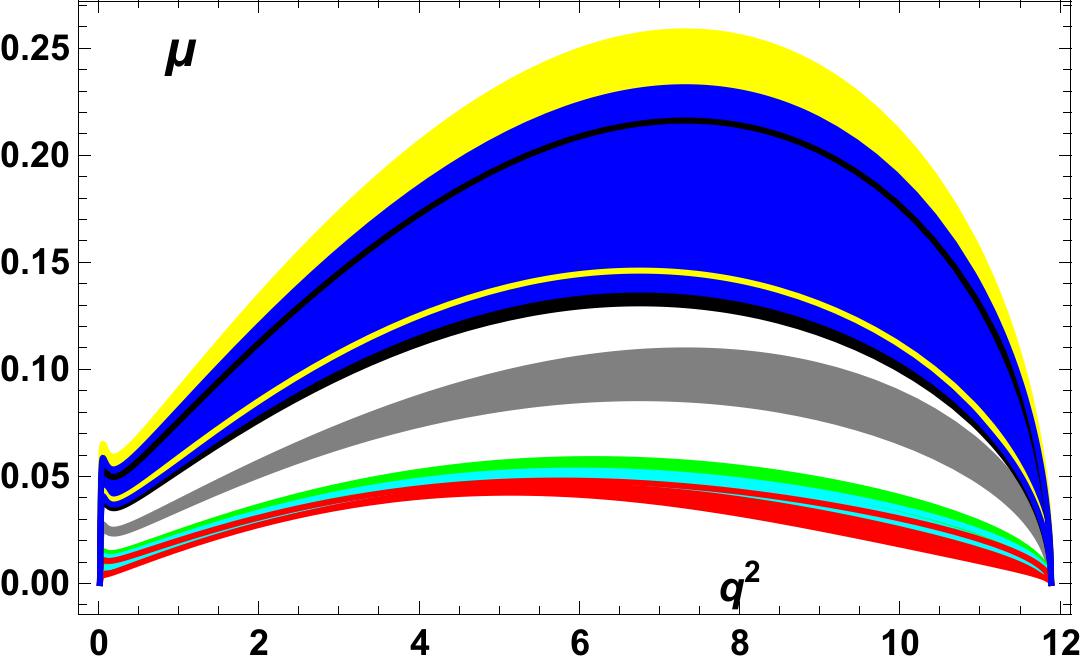} 
\includegraphics[width=0.3\textwidth]{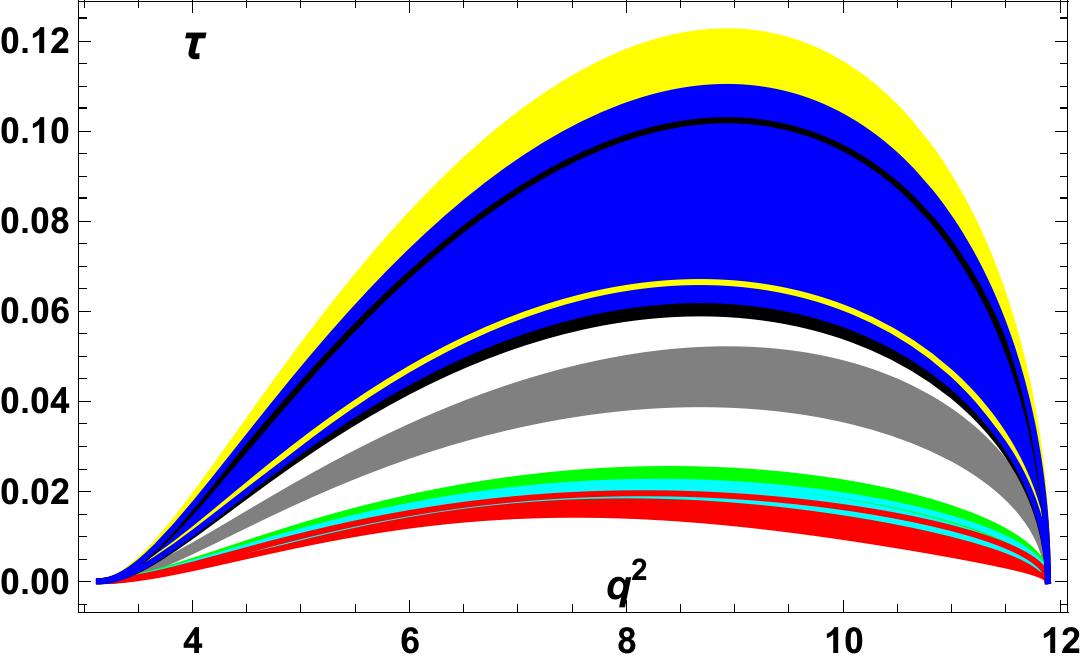}\\
\includegraphics[width=0.3\textwidth]{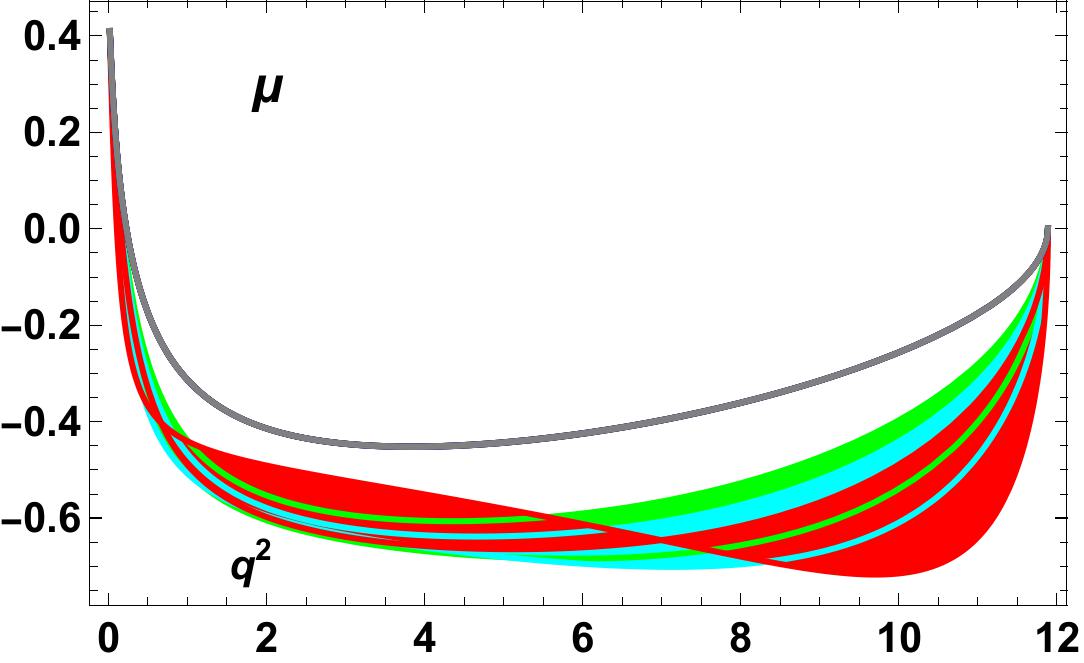}
\includegraphics[width=0.3\textwidth]{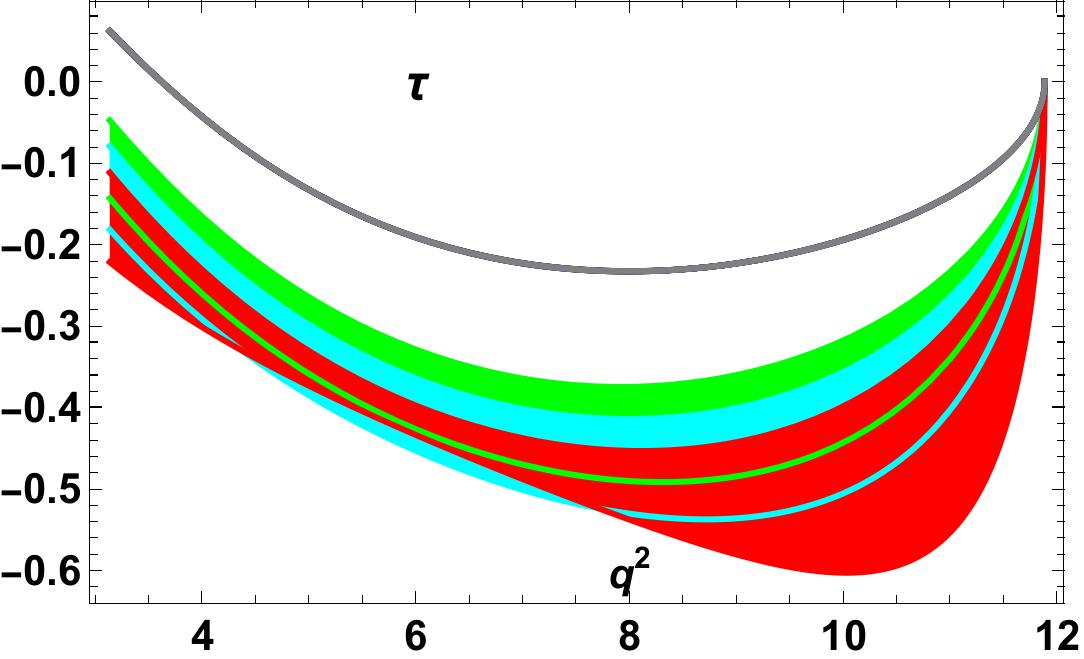}
\caption{$\text{Br}\times 10^8$ (first two plots) and $A_{FB}$ (last two plots) of  $B^{*0}_{s} \rightarrow 
\left[
   D_s^-(\rightarrow \tau^-\,\bar\nu_{\tau})
\right]\,
\ell^{+}\,{\nu}_\ell$ for decay to the muon and the tauon.}
\label{Branching Vs q2 BsDs}
\end{figure}

In Figure~\ref{Branching Vs q2 BsDs}, we present the result for the branching ratio (first row) and forward-backward asymmetry 
(second) as a function of the dilepton momentum transfer squared ($q^2$) for the 
$B^{*0}_{s} \rightarrow 
\left[
   D_s^-(\rightarrow \tau^-\,\bar\nu_{\tau})
\right]\,
\ell^{+}\,{\nu}_\ell$ decay in both the case of the muon (first column) and tauon (second column) final states. 
The gray color in this figure shows the SM result with the width corresponding to the uncertainty in the form factors. 
We show the benchmark points BI-BVI of Table~\ref{benchmark} using the colors black (BI), green (BII), blue (BIII), cyan (BIV), yellow (BV), and red (BVI). The width of these bands corresponds to the uncertainties arising from the form factors as well as the $1\sigma$ 
uncertainty in the corresponding WC. 
If we focus on the first row of Figure~\ref{Branching Vs q2 BsDs}, we can see that the branching fraction is primarily an increasing (decreasing) function of $q^2$ for $q^2\lesssim7$ ($q^2\gtrsim7$). It is also noticeable that each NP bench mark can significantly modify the decay rate. We further note that for the NP bench marks considered in this study, a positive left-handed coupling $C_{V_L}$ (black~[BI], blue~[BIII], and yellow~[BV]) constructively interferes with the SM current, raising the branching fraction above the SM value (gray band) throughout the $q^2$ region. On the other hand, the benchmarks corresponding to the right-handed coupling $C_{V_R}$ (green~[BII], cyan~[BIV], and red~[BVI]) 
lower the branching fraction value throughout the $q^2$ region, compared with the SM result. The most significant deviations are seen in benchmarks BV and BVI, which have the highest and lowest branching fraction values, respectively. As the branching ratio is sensitive to both left and right-handed currents, precision measurements the differential branching ratio could constrain $C_{V_L}$ and $C_{V_R}$. In particular, any appreciable deviation from the SM expectation can signal NP in either chirality and complement the existing constraints from $b\to c\tau\nu$ observables. Whereas the $C_{V_L}>0$ and $C_{V_R}>0$ contribute with opposite signs to the branching ratio, the effects of different benchmark points for $C_{V_L}$ and $C_{V_R}$ exhibit significant overlap. This makes it challenging to distinguish between the benchmark points based on, for example, $C_{V_L}>0$. Consequently, it is of interest to explore whether other observables can more effectively differentiate these benchmark points in the presence of NP contributions. 

In the second row of Figure~\ref{Branching Vs q2 BsDs}, we plot the $A_{FB}$ as a function of $q^2$. It is interesting to note that the $A_{{FB}}$ is not at all sensitive to $C_{V_L}$ as it shows no deviation from the SM value. On the other hand $A_FB$ is quite sensitive to the \( C_{V_R}\) NP and have a visible deviations from their SM values. For instance, all of the benchmarks corresponding to \( C_{V_R}\) that we consider enhance the probability of the final state lepton in the backward direction throughout the $q^2$ region. The maximum shift from the SM in the value of $A_{FB}$ around $q^2\simeq10$ GeV$^2$ due to the BII, BIV and BVI is, respectively, about $24\%$ ($33\%$), $38\%$ ($50\%$) and $66\%$ ($92\%$), for the case of $\mu$ ($\tau$).

\begin{figure}[H]
 \centering
\includegraphics[width=1.5in]{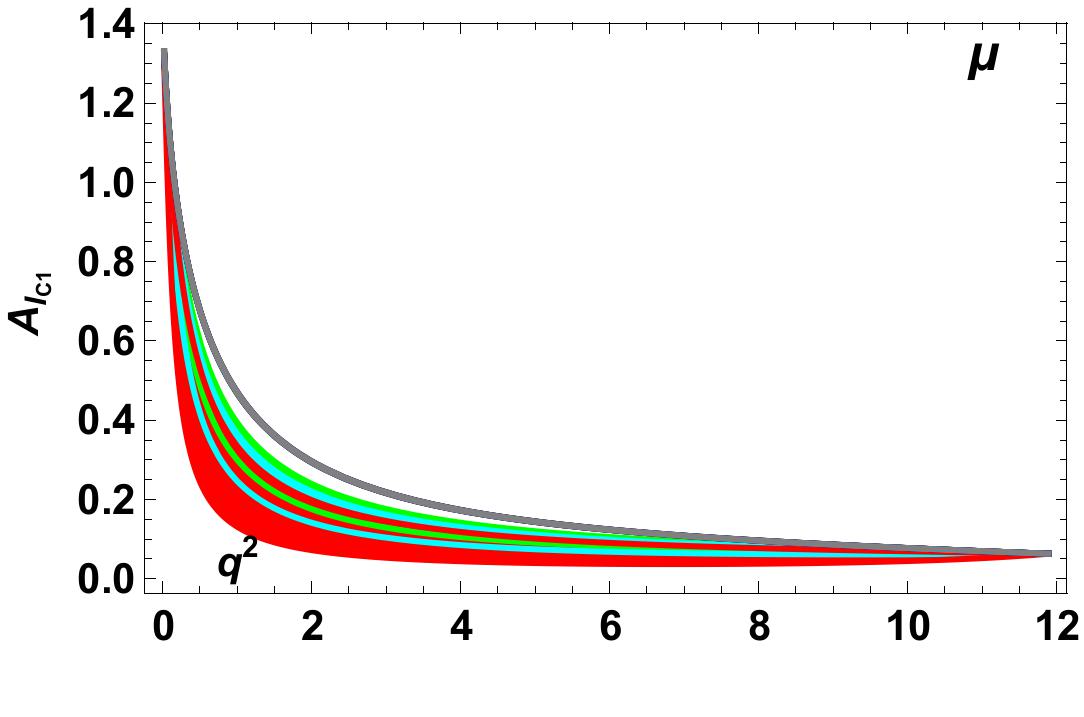}
\includegraphics[width=1.5in]{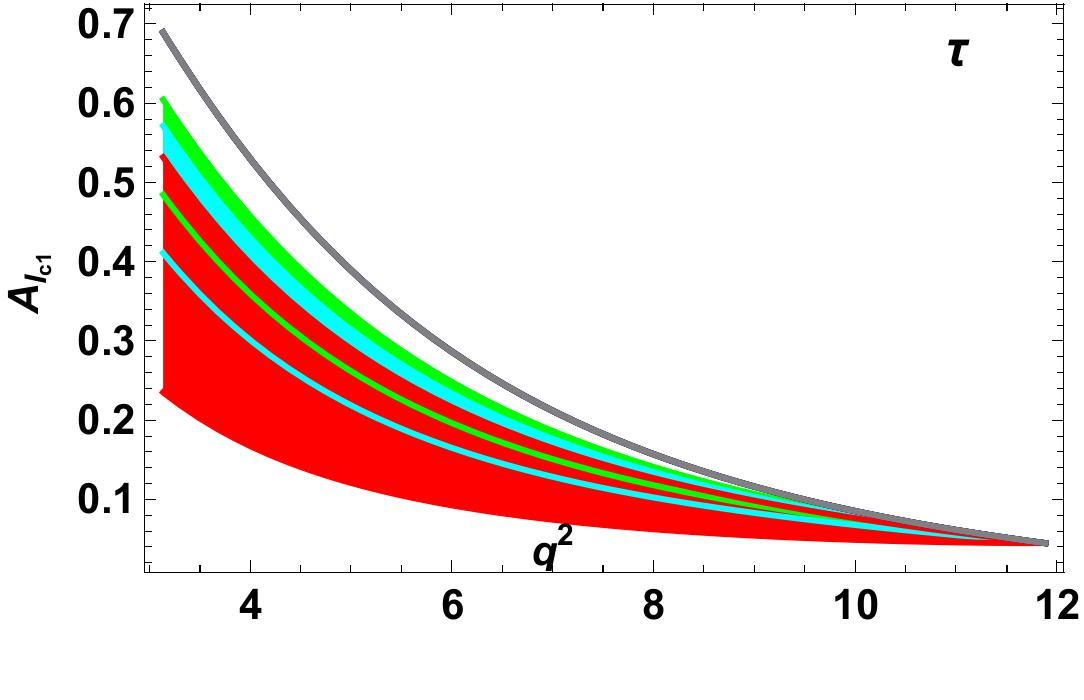}
\includegraphics[width=1.5in]{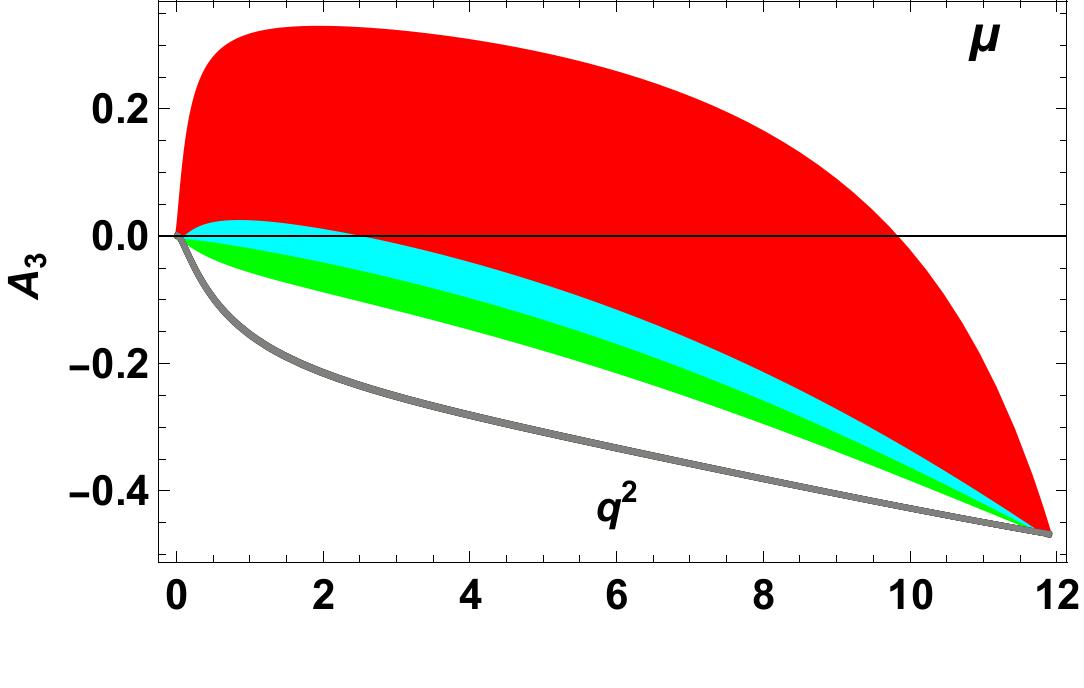}
\includegraphics[width=1.5in]{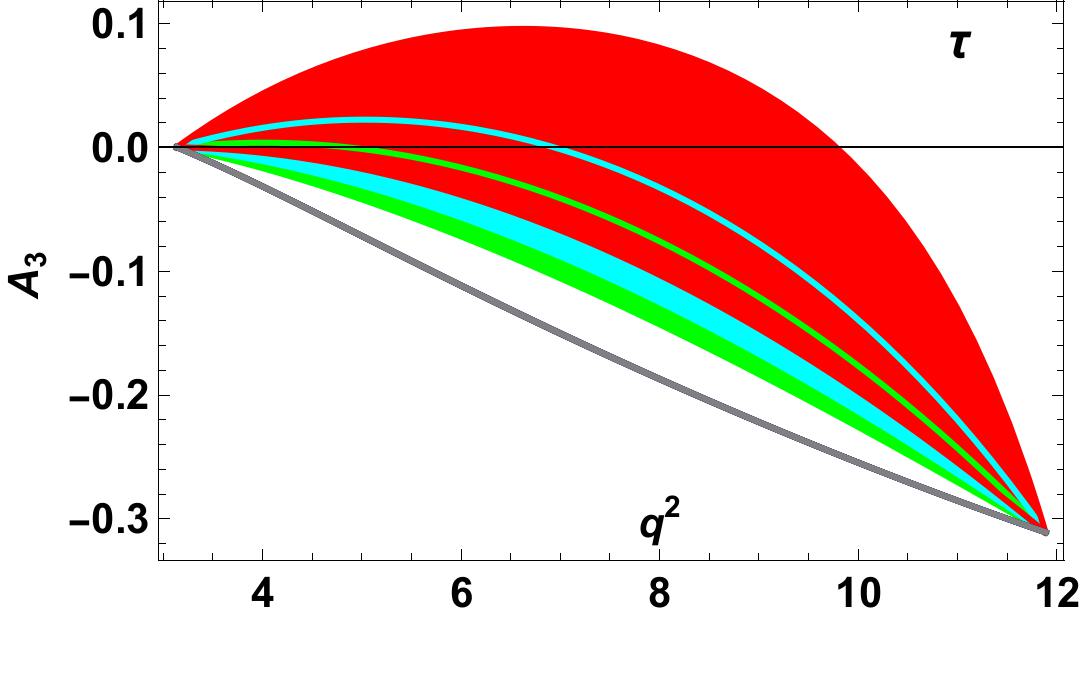}
\includegraphics[width=1.5in]{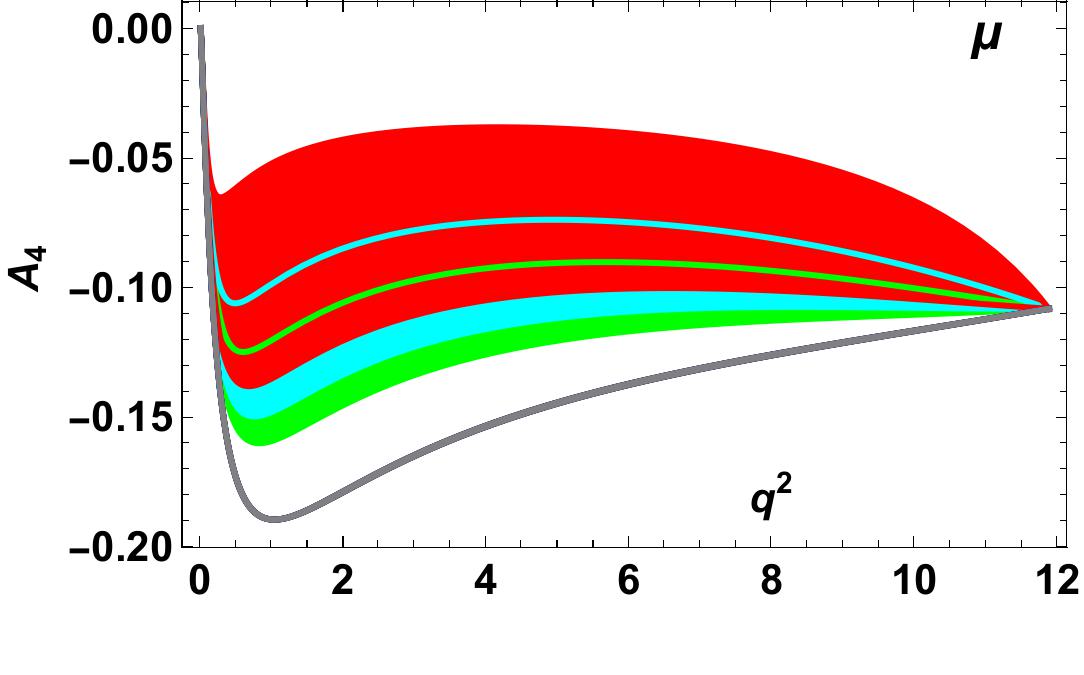}
\includegraphics[width=1.5in]{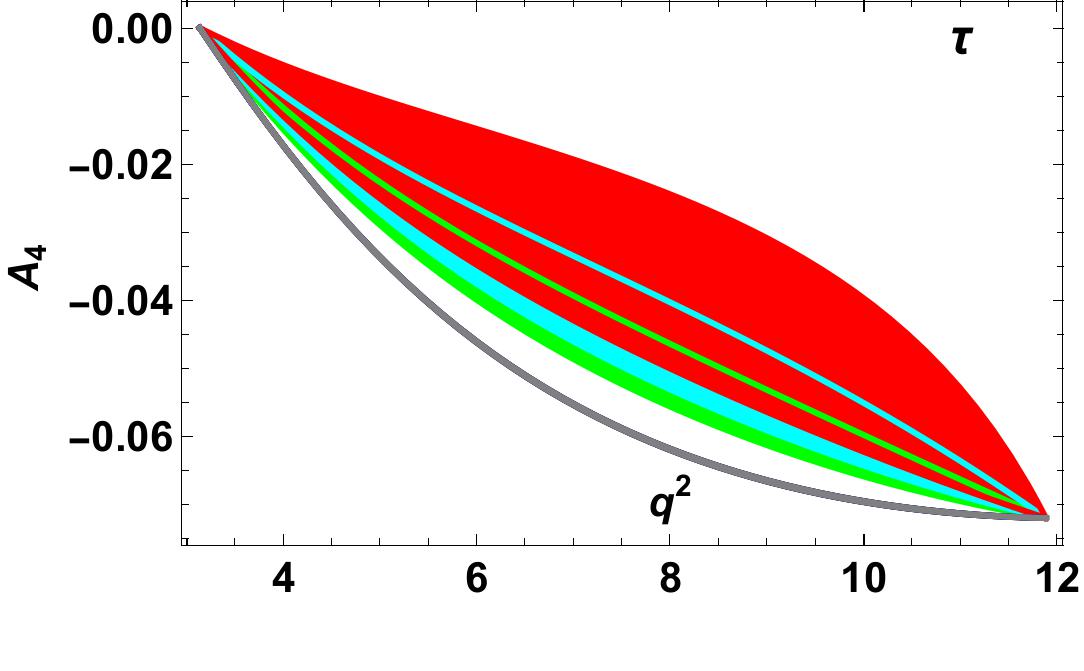}
\includegraphics[width=1.5in]{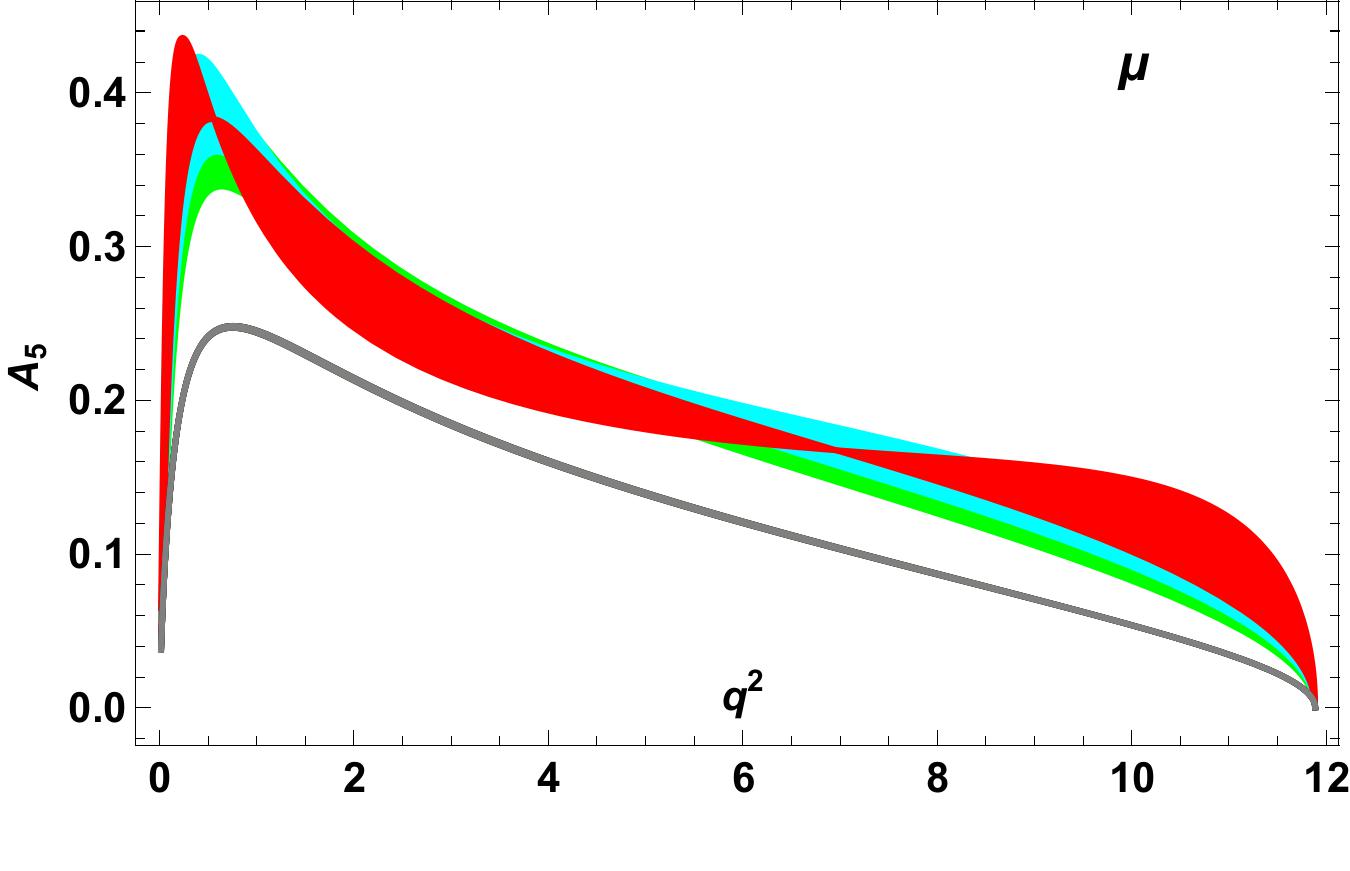}
\includegraphics[width=1.5in]{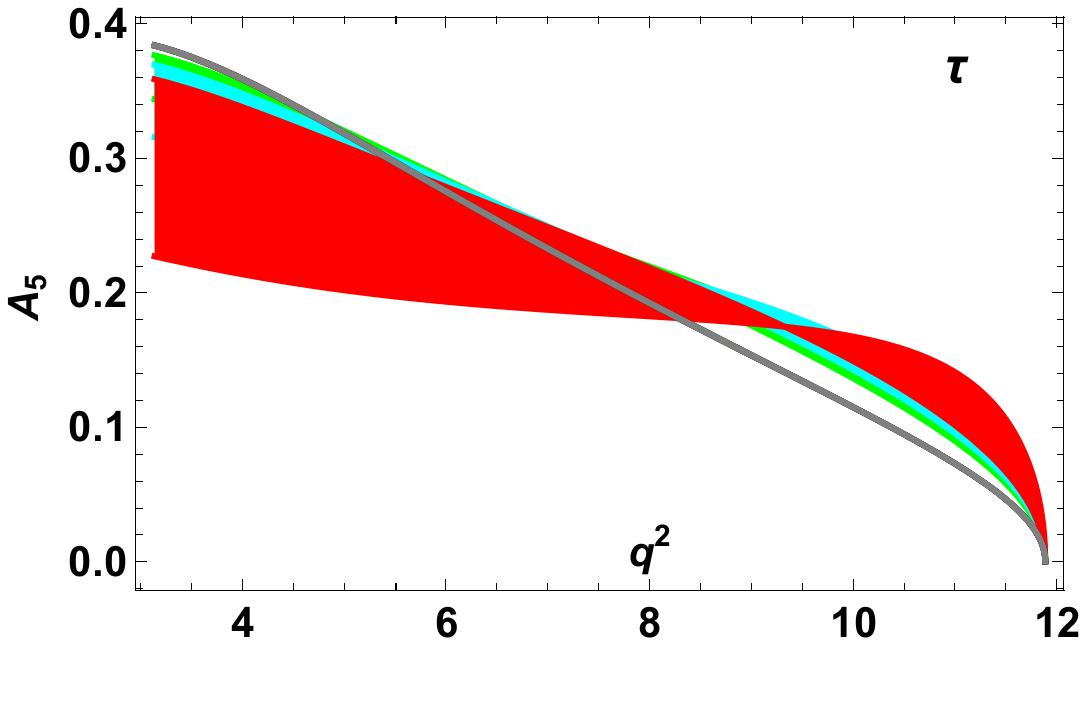}
\includegraphics[width=1.5in]{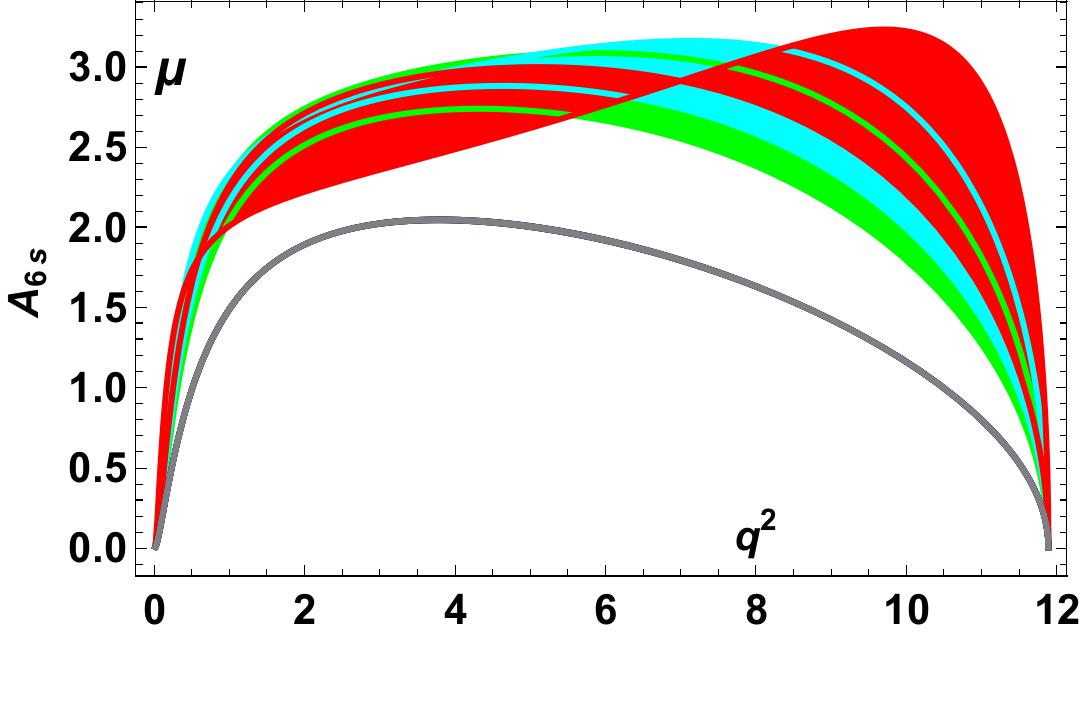}
\includegraphics[width=1.5in]{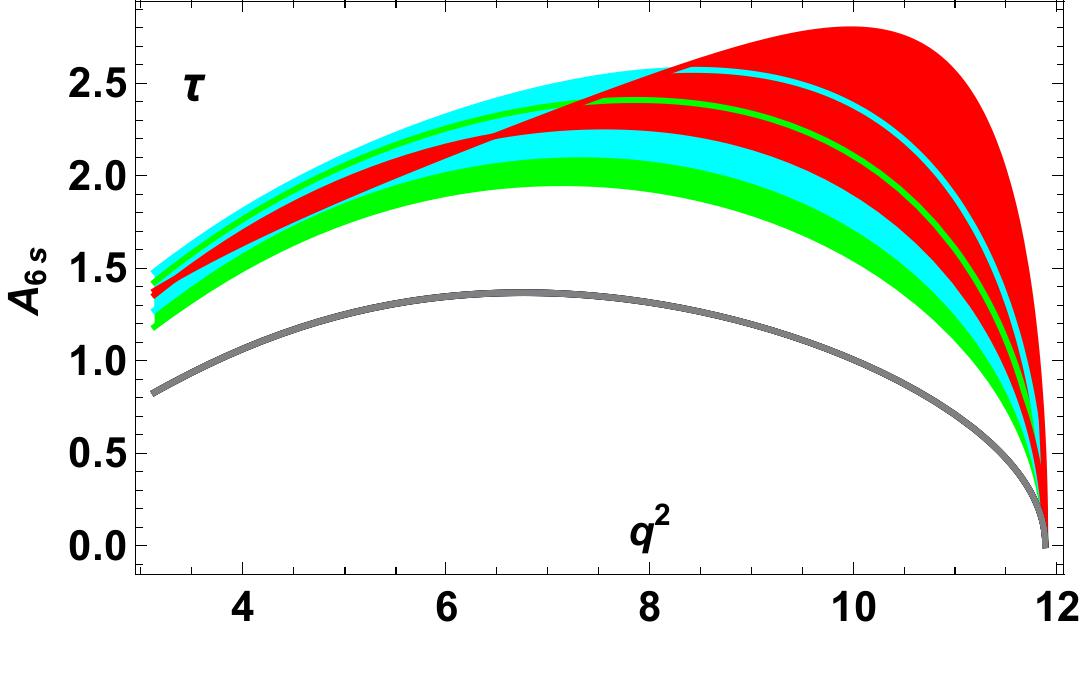}
\caption{Angular Observables $A_{I_{c1}}$, $A_3$, $A_4$, $A_5$, and $A_{6s}$  for  muon and tauon for $B^{*0}_{s} \rightarrow 
\bigl[
   D_s^-(\rightarrow \tau^-\,\bar\nu_{\tau})
\bigr]\,
\ell^{+}\,{\nu}_\ell$ decay.}
\label{angular observables of bs to ds}
\end{figure}

A similar trend can be seen in Figure~\ref{angular observables of bs to ds} where we plot the $q^2$ dependence of the angular observables $A_i$ ($i=3,4,5,6s, I_{c1}$) for both the $\mu$ and $\tau$ cases. These angular observables, like the $A_{FB}$, are insensitive to $C_{V_L}$ but do show significant deviation from the SM result in the case of $C_{V_R}$ NP (BII, BIV, and BVI). In particular, BVI (red band) shows the widest uncertainty band due to large form factor uncertainties throughout the kinematical region, and, therefore shows the most prominent shift from the SM. We do note that notwithstanding the overlap among the uncertainty bands of different benchmarks, these observables provide valuable insights. 

The first two plots in the first row of Figure~\ref{angular observables of bs to ds} depict the $q^2$ dependence of $A_{I_{c1}}$ for $\mu$ and $\tau$ where the effects of NP are prominent at low $q^2$, particularly, for the case of $\tau$. The last two plots of the first row of Figure~\ref{angular observables of bs to ds} show $A_3$ for $\mu$ and $\tau$. In the SM and in the case of BII, the value of this observable for both $\mu$ and $\tau$ cases is -ve throughout the kinematical region while benchmarks BIV and BVI can shift its value to the +ve with the largest shift occuring due to the BVI. Furthermore, in benchmarks BIV and BVI, this observable has a zero crossing, which in the case of BIV interestingly occurs in the low $q^2$ region. The zero crossing and the different signature in the values of this observable may be useful for not only putting more constraints on the NP parametric space but also for distinguishing these benchmarks. 

In the second row of Figure~\ref{angular observables of bs to ds} we show the result of $A_4$ in both the $\mu$ and $\tau$ cases as the first two plots. 
In the case of the $\mu$, $A_4$ peaks towards the middle of the $q^2$ region.  As expected, the largest shift in its peak occurs for the BVI benchmark owing to large uncertainties. It is also interesting to note the behavior of the minimum in the value of 
$A_4$ (for the $\mu$ case) the location of which shifts to a lower $q^2$ value. We can finally note that $A_4$ receives +ve contributions in both 
the $\mu$ and $\tau$ cases over the entire kinematical range. 

In the last two plots of the second row of Figure~\ref{angular observables of bs to ds}, we present the result of $A_5$ for both the 
$\mu$ and $\tau$ cases. For the case of the $\mu$, $A_5$ displays a sharp peak in the low $q^2$ region $(\sim 1-3)$GeV$^2$, which is 
most pronounced for the BVI benchmark, as expected. The contribution to $A_5$ over the entire kinematical range is +ve in the 
case of the $\mu$. However, in the case of the $\tau$, the NP contribution to $A_5$ changes signature from -ve to +ve 
in the mid $q^2$ ($6-8$GeV$^2$) region.    

\begin{figure}[H]
\centering
\includegraphics[width=3in,height=2in]{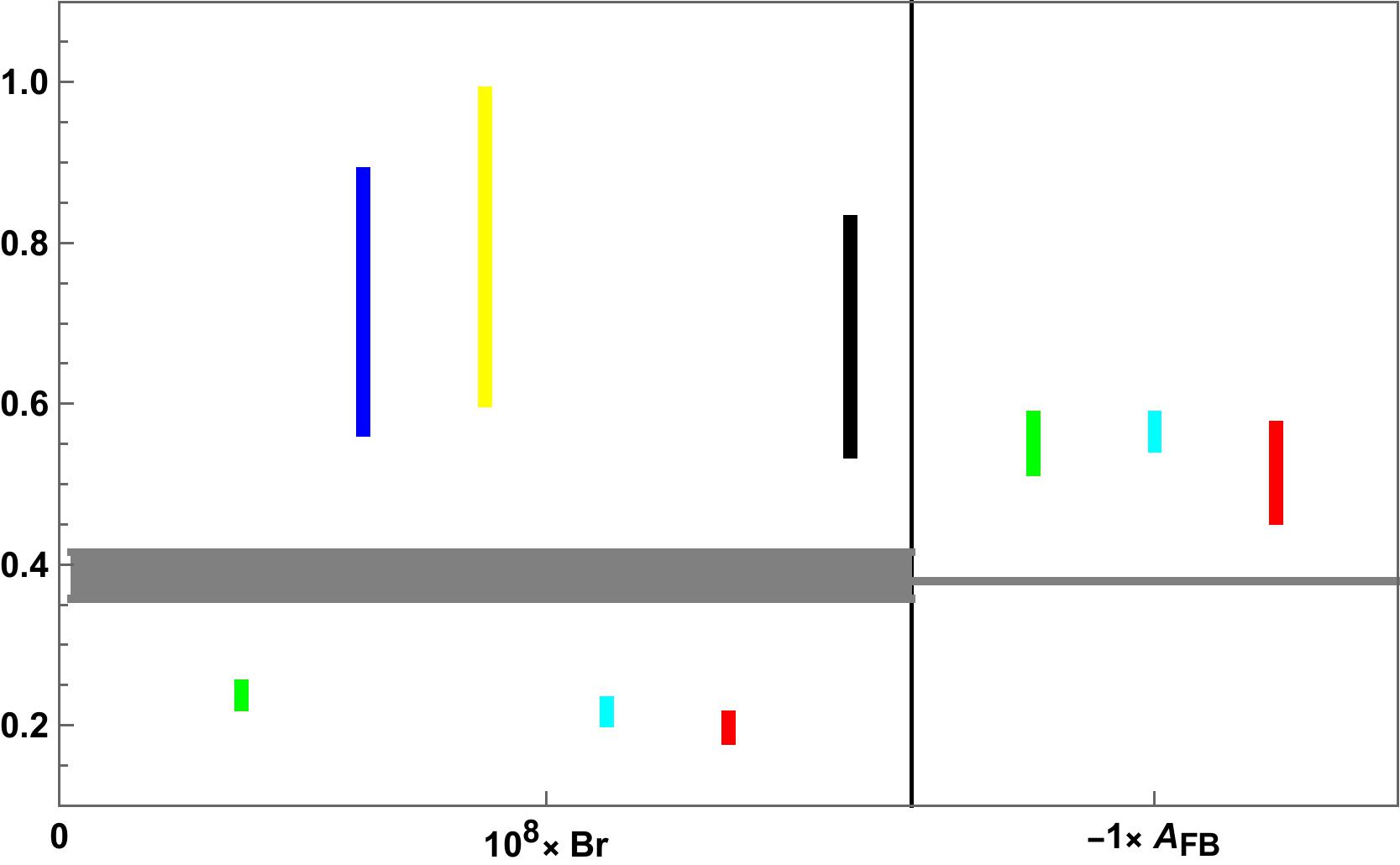}
\includegraphics[width=3in,height=2in]{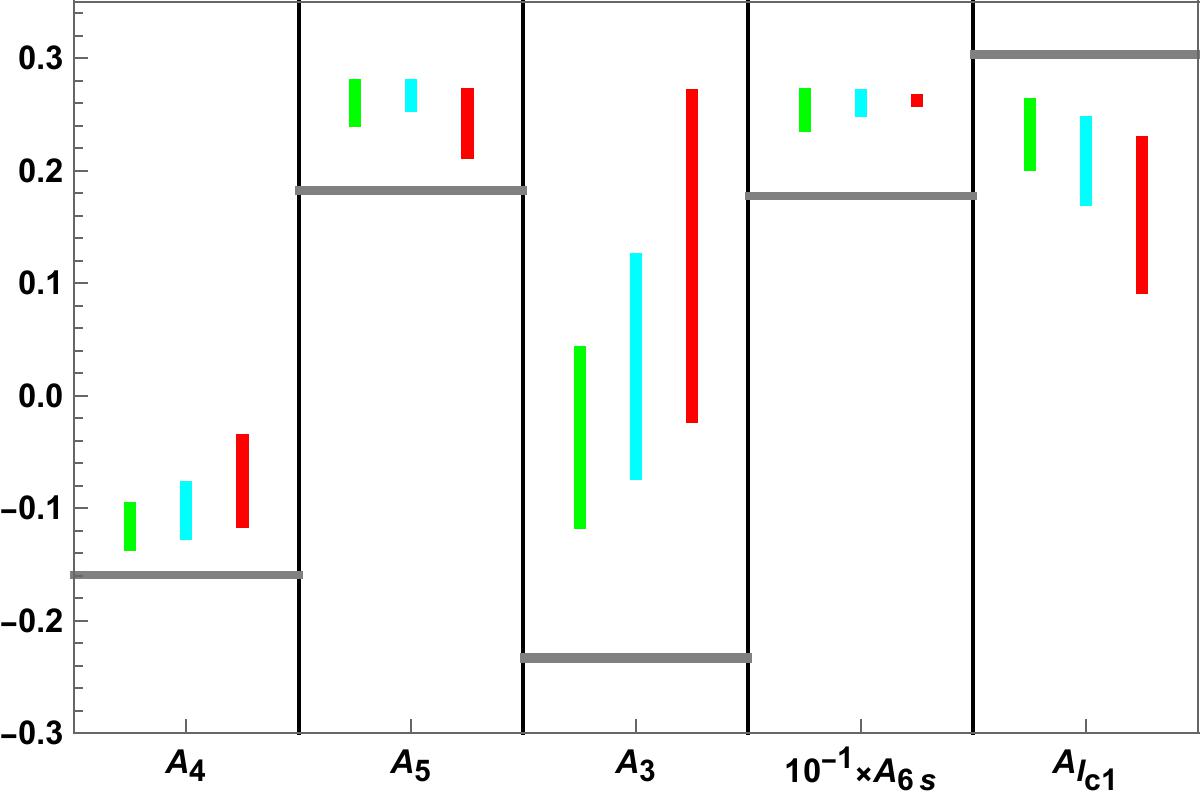}
\includegraphics[width=3in,height=2in]{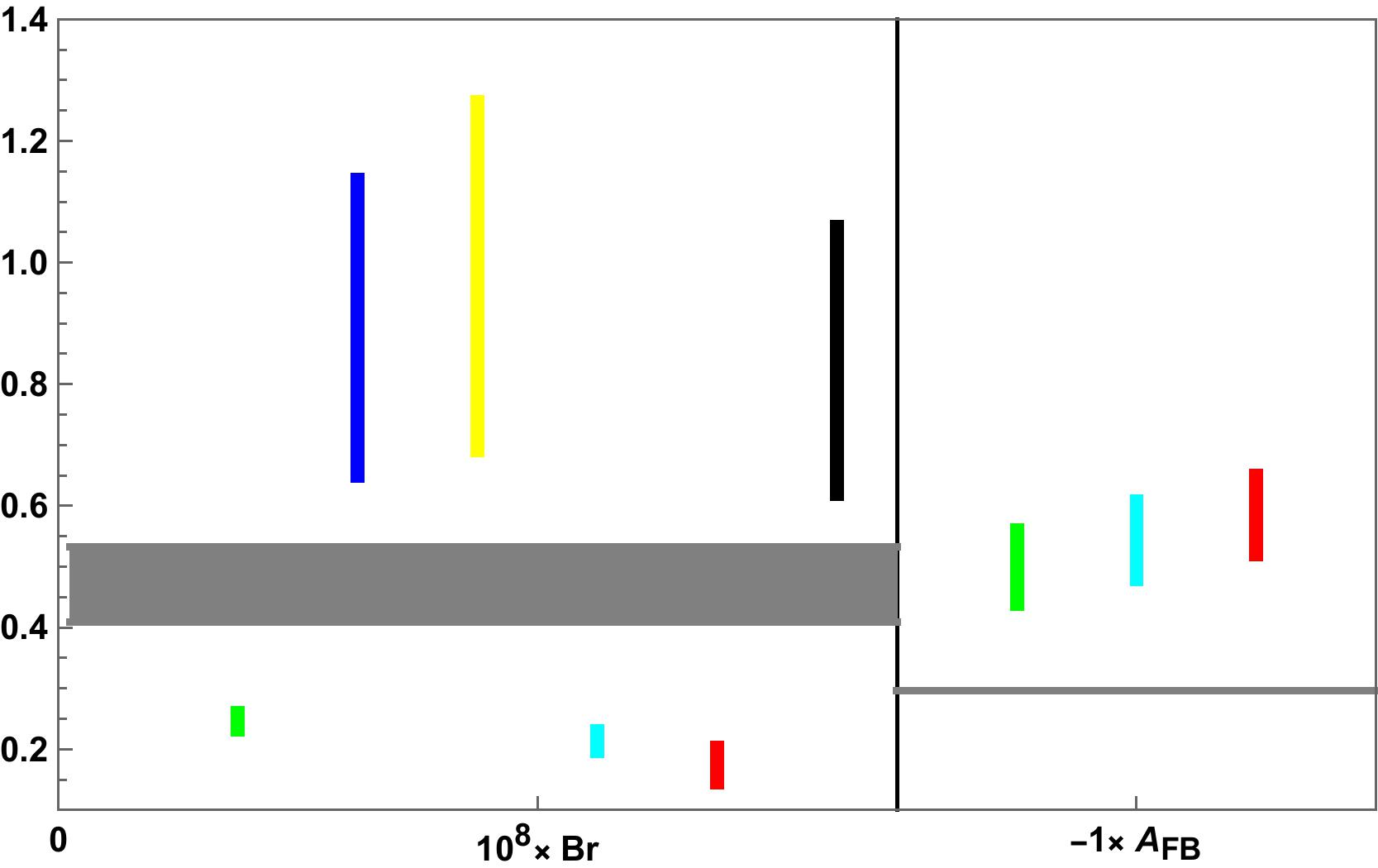}
\includegraphics[width=3in,height=2in]{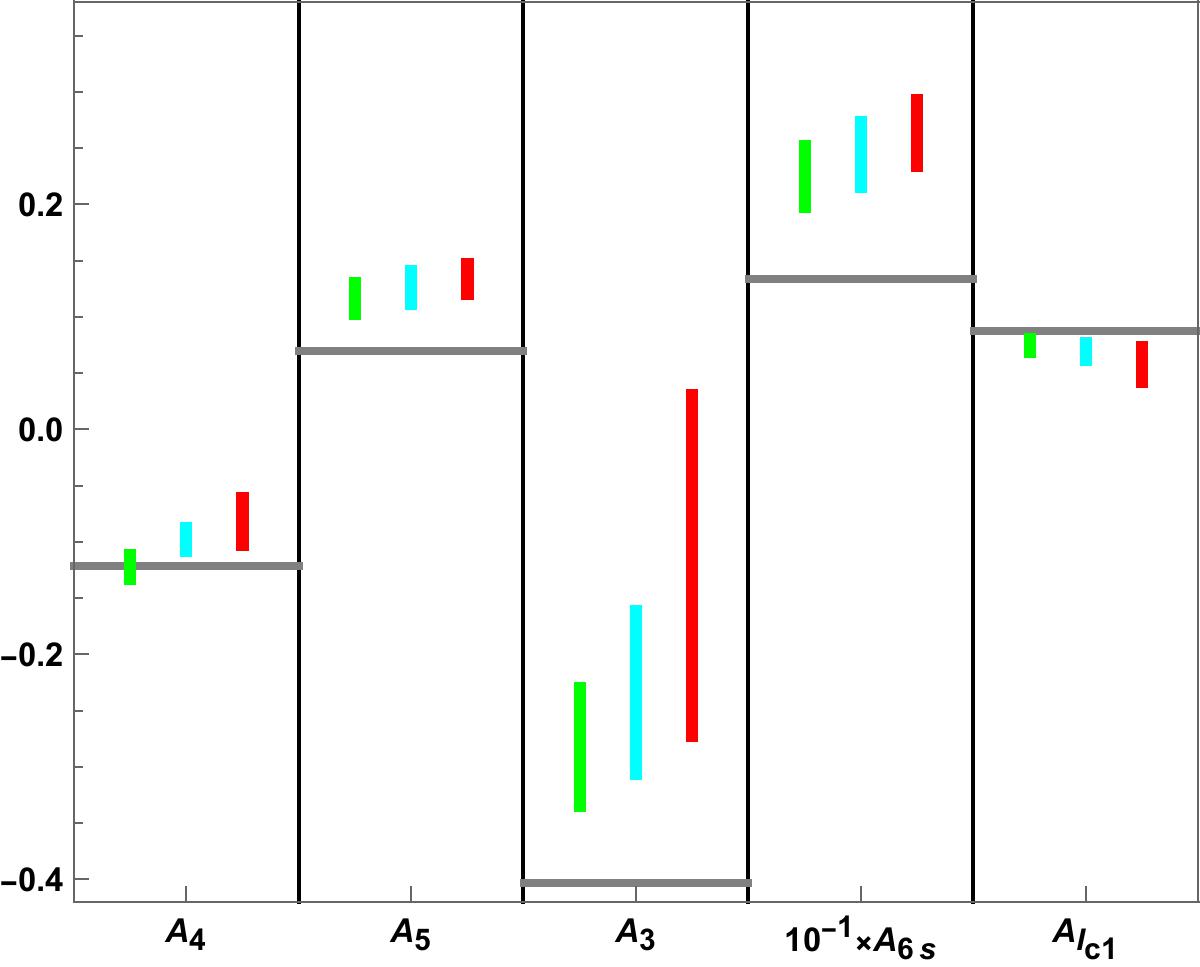}
\caption{The variation in the magnitudes of Br, $A_{FB}$, and angular observables $A_i$ due to the presence of NP. The first two plots correspond to low $q^2\in[s_{\text{min}},6]$GeV$^2$ and the last two plots correspond to high $q^2\in[6,s_{\text{max}}]$GeV$^2$ for  $B^{*0}_{s} \rightarrow 
\bigl[
   D_s^-(\rightarrow \tau^-\,\bar\nu_{\tau})
\bigr]\,
\mu^{+}\,{\nu}_\mu$ decay.}
\label{lowmu}
\end{figure}

\begin{figure}[H]
\centering
\includegraphics[width=3in,height=2in]{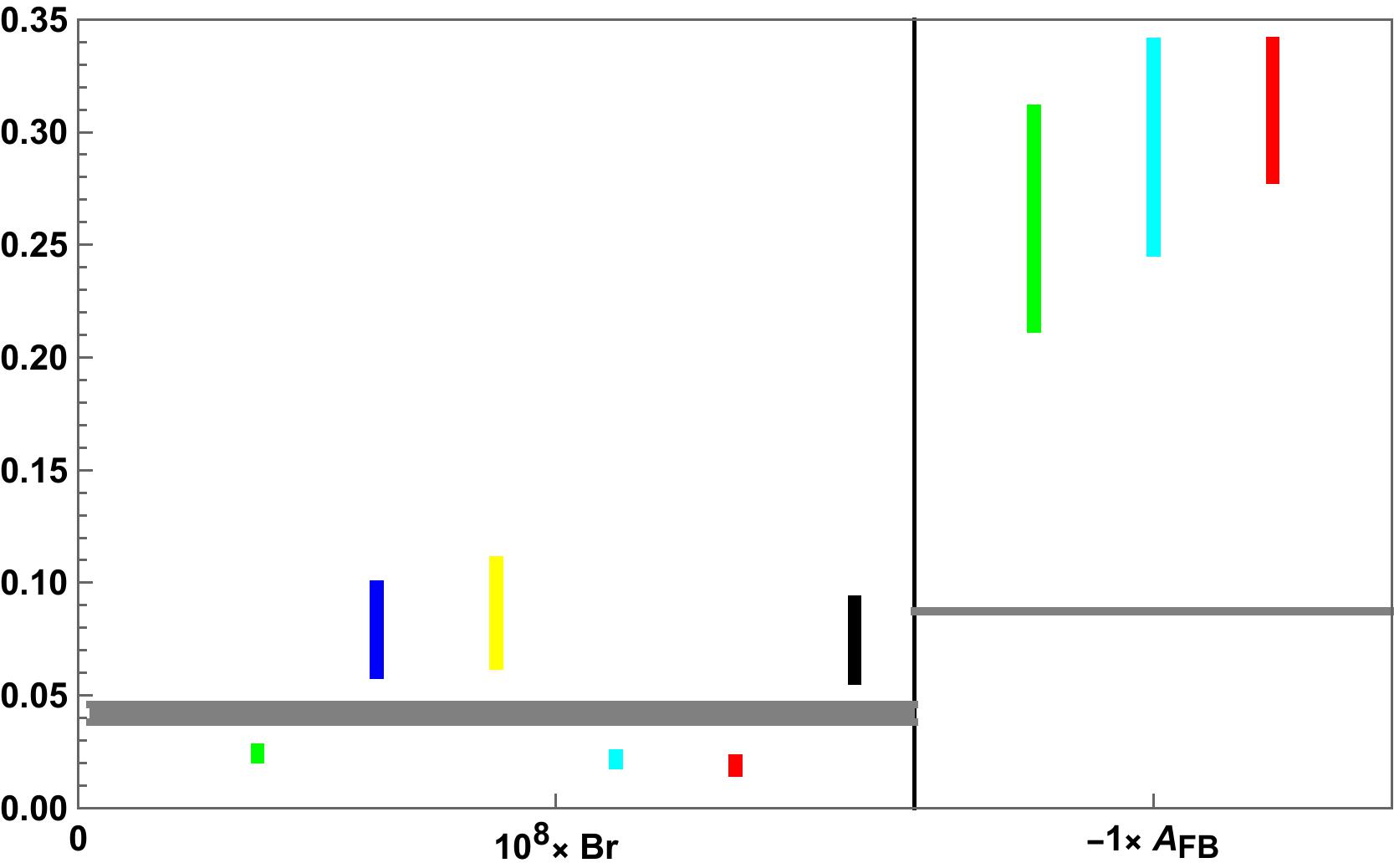}
\includegraphics[width=3in,height=2in]{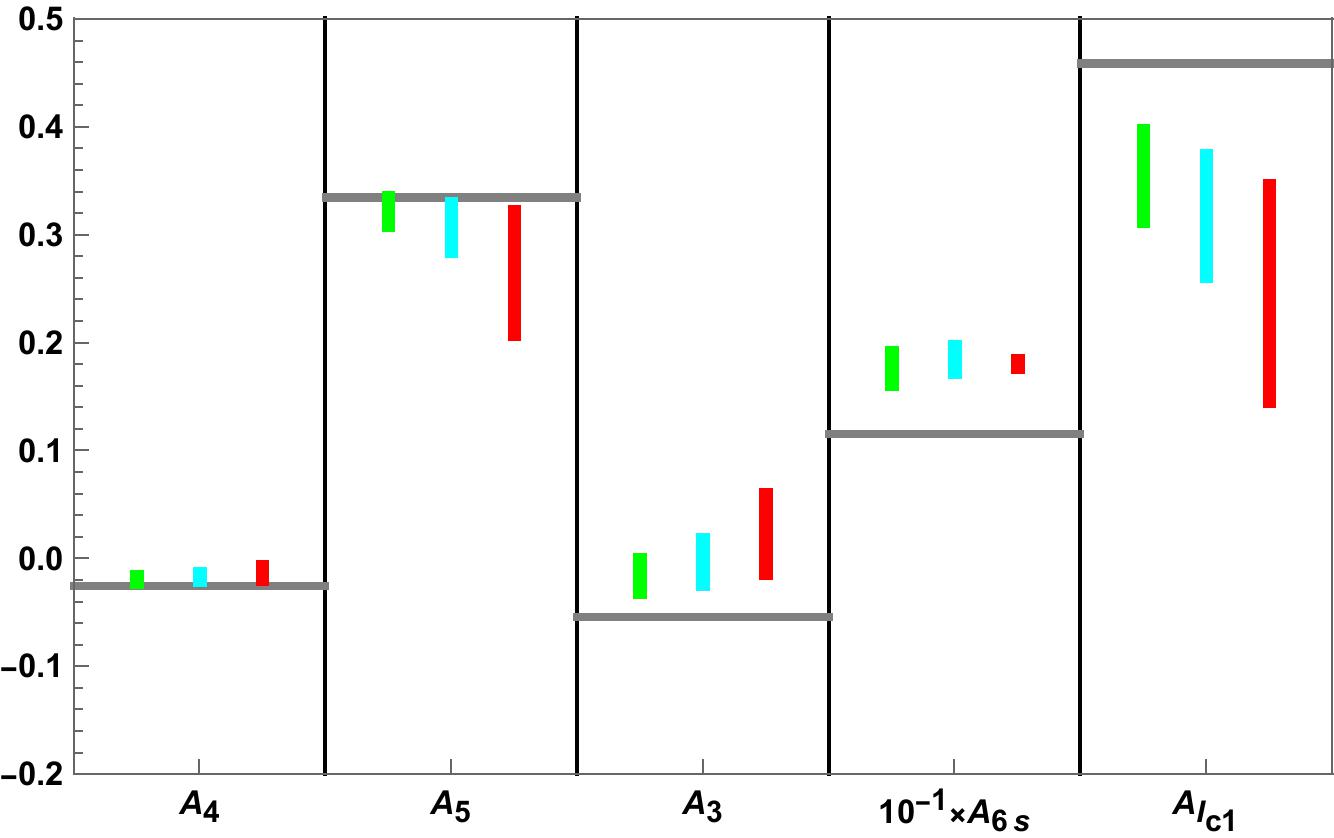}
\includegraphics[width=3in,height=2in]{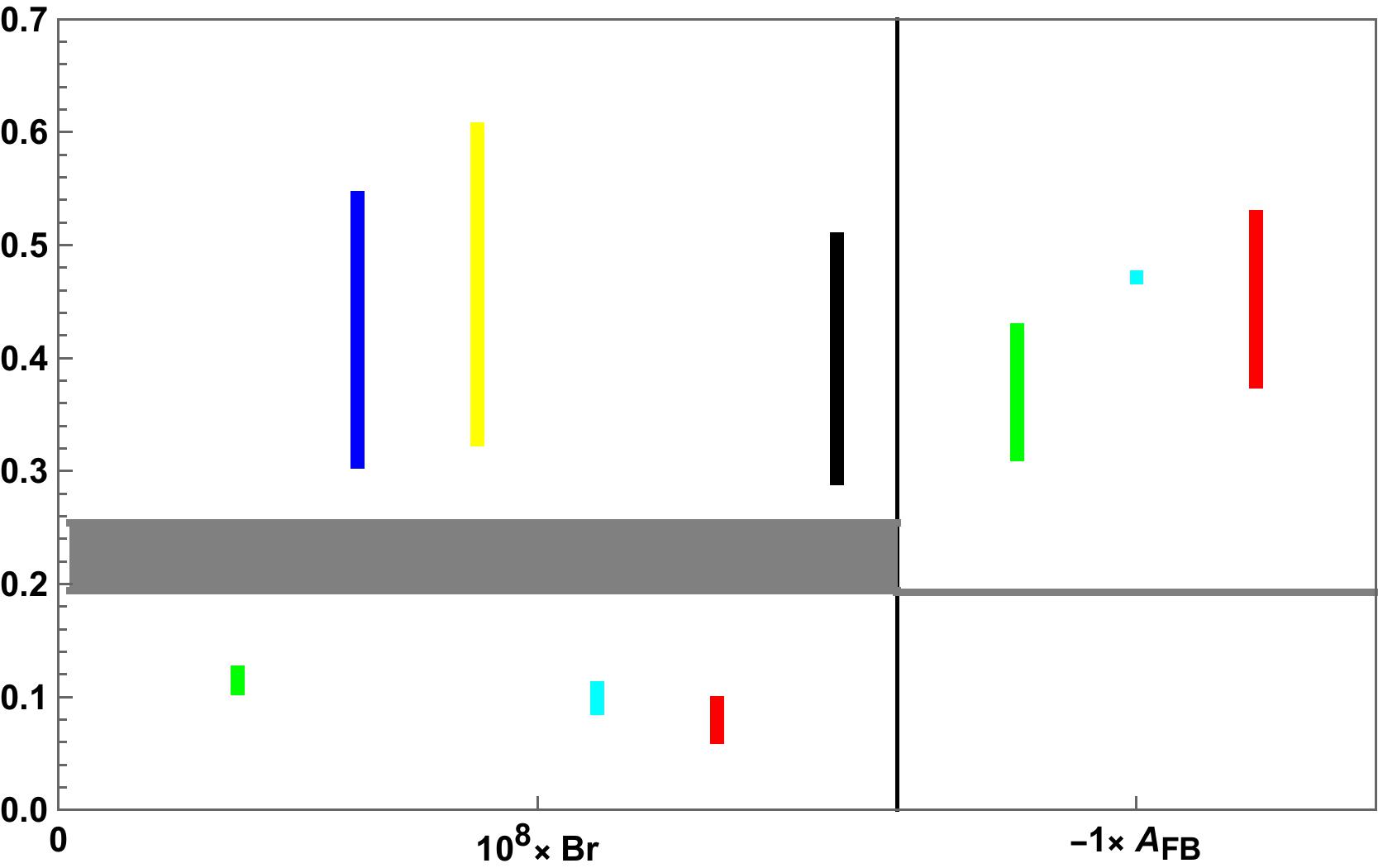}
\includegraphics[width=3in,height=2in]{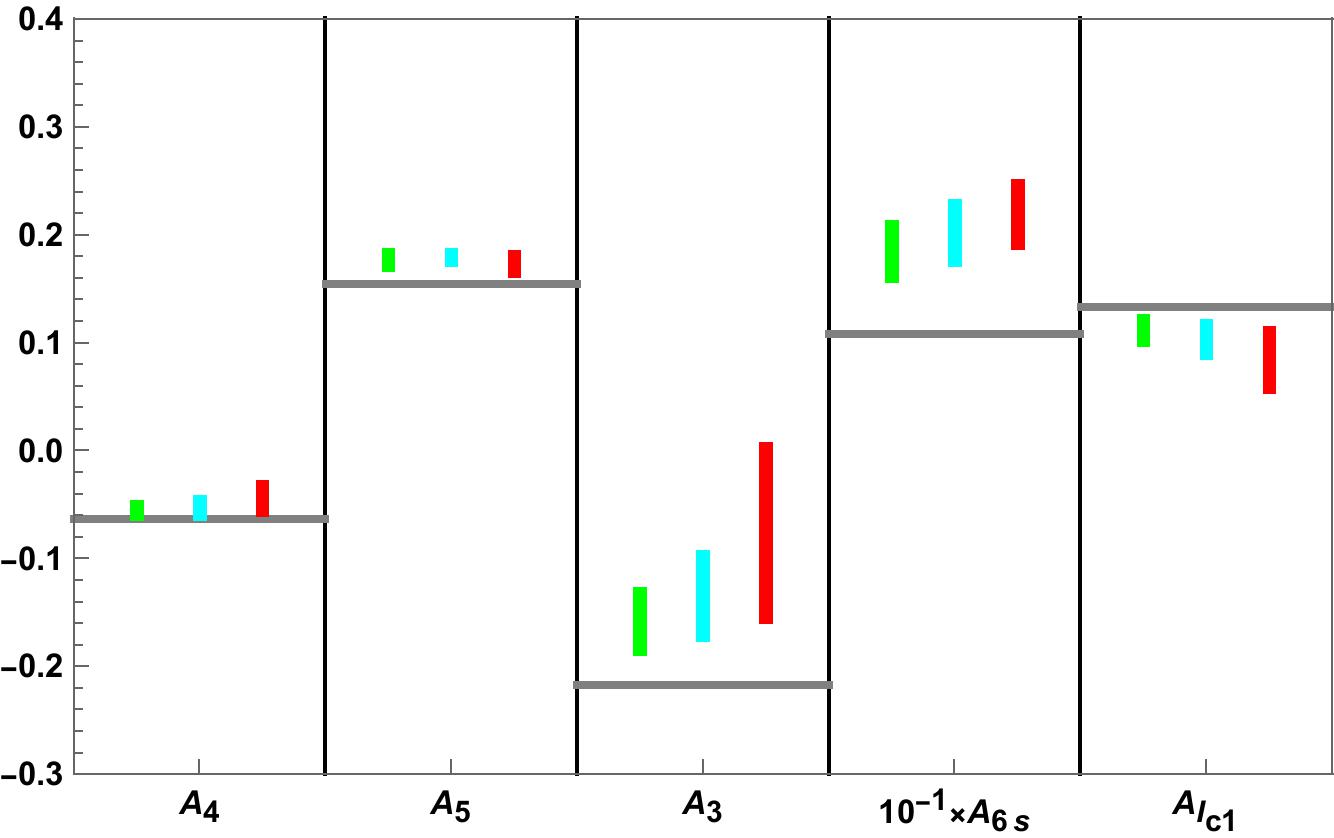}
\caption{The variation in the magnitudes of Br, $A_{FB}$, and angular observables $A_i$ due to the presence of NP. The first two plots correspond to low $q^2\in[s_{\text{min}},6]$GeV$^2$ and the last two plots correspond to high $q^2\in[6,s_{\text{max}}]$GeV$^2$ for  $B^{*0}_{s} \rightarrow 
\bigl[
   D_s^-(\rightarrow \tau^-\,\bar\nu_{\tau})
\bigr]\,
\tau^{+}\,{\nu}_\tau$ decay.}
\label{lowtau}
\end{figure}

In the third row of Figure~\ref{angular observables of bs to ds} we show the result for $A_{6s}$ for the $\mu$ and $\tau$ cases. 
These plots show appreciable deviations from the SM in both the $\mu$ and $\tau$ cases over the entire $q^2$ region. In particular, $A_{6s}$ peaks 
towards the high $q^2$ region which for benchmark BVI shows a maximum at $\sim 10$GeV$^2$ for the case of the $\mu$ and at $\sim 11$GeV$^2$ for the case of the $\tau$. The other benchmarks BII and BIV display  a maximum in $q^2$ at a slightly lower value but still in the high $q^2$ region.  
It is interesting to look at the physical observables after integration over low and high $q^2$ regions. In Figure~\ref{lowmu} (Figure~\ref{lowtau}) we show this 
result for the decay to the $\mu$ ($\tau$), with the first row giving the integrated bar plots for the low $q^2\in (s_{min}, 6~{\rm GeV^2})$ bin, where $s_{min}$ value can be $m_\mu^2$ or $m_\tau^2$. 
The second row giving the corresponding results for the high $q^2\in (6~{\rm GeV^2}, s_{max}\equiv 11.8~{\rm GeV^2})$ bin. Consistent with the trend seen in Figure~\ref{Branching Vs q2 BsDs}, we can see in both Figure~\ref{lowmu} and Figure~\ref{lowtau} that $C_{V_L}$ type NP increases the branching fraction of the decay while $C_{V_{R}}$ reduces. Whereas the $C_{V_{L}}$ and $C_{V_R}$ 
benchmarks can be clearly distinguished from each other based on the Br, it is clear that the 
three benchmarks within any one of these scenarios overlap with each other. 

We can again see in Figure~\ref{lowmu} and Figure~\ref{lowtau} that the forward-backward asymmetry and the angular observables are 
not sensitive to $C_{V_L}$ NP which predicts a result consistent with the SM. At the same time, $C_{V_R}$ does contribute NP effects to these observables. 
The reason for this behavior lies in the angular dependence of the decay amplitude. The forward-backward asymmetry is influenced by the interference of different helicity amplitudes. In particular, it seems that the right-handed couplings can alter the angular distribution of the final-state leptons in a manner not possible with purely left-handed interactions. Hence, even small values of \(\displaystyle C_{V_R}\) can produce noticeable shifts in \(A_{\mathrm{FB}}\), whereas \(\displaystyle C_{V_L}\) contributions often mimic the SM structure more closely. We discuss this point further in Section~\ref{cTos_transition} where we show that this behavior 
extends to other charged current processes as well. 

Focusing for now the $C_{V_R}$ NP contributions to the forward-backward asymmetry and 
the angular observables, we can see that in most cases the benchmark points discussed do overlap with each other. Having said this, 
it is clear that these benchmarks are quite well separated from the SM result. Further, we can see, for example in the bar plot for $A_3$, BII and BVI can potentially lead to results that may very well be different from each other. 

The numerical values of all observables are listed in Tables \ref{tab:bs_decay_mu_low}, \ref{tab:bs_decay_mu_high}, \ref{tab:bs_decay_tau_low} and \ref{tab:bs_decay_tau_high}. In particular, we give values for the case of the $\mu$ in 
the low (high) $q^2$ bin in Table~\ref{tab:bs_decay_mu_low} (\ref{tab:bs_decay_mu_high}). The corresponding values for 
the low (high) $q^2$ bin are given in Table~\ref{tab:bs_decay_tau_low} (\ref{tab:bs_decay_tau_high}) for the case of the 
$\tau$. 

 \begin{figure}[H]
 \centering
 \includegraphics[width=2in]{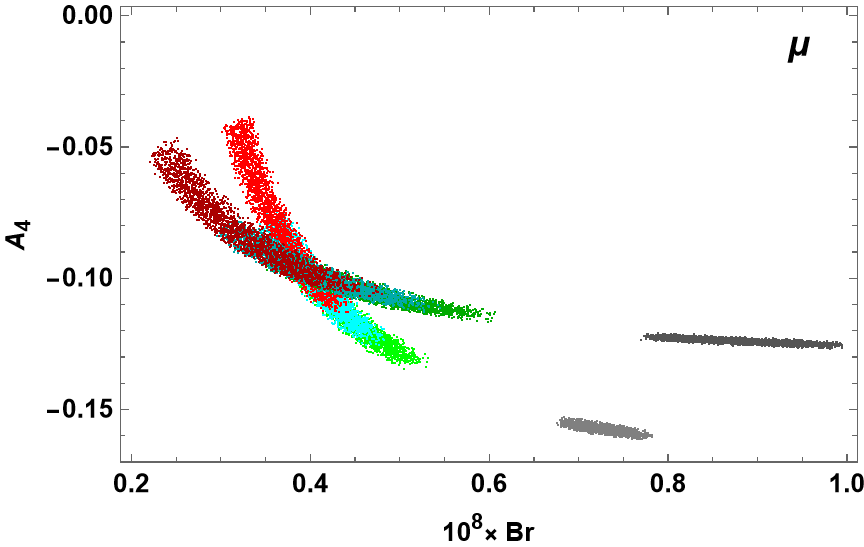}
 \includegraphics[width=2in]{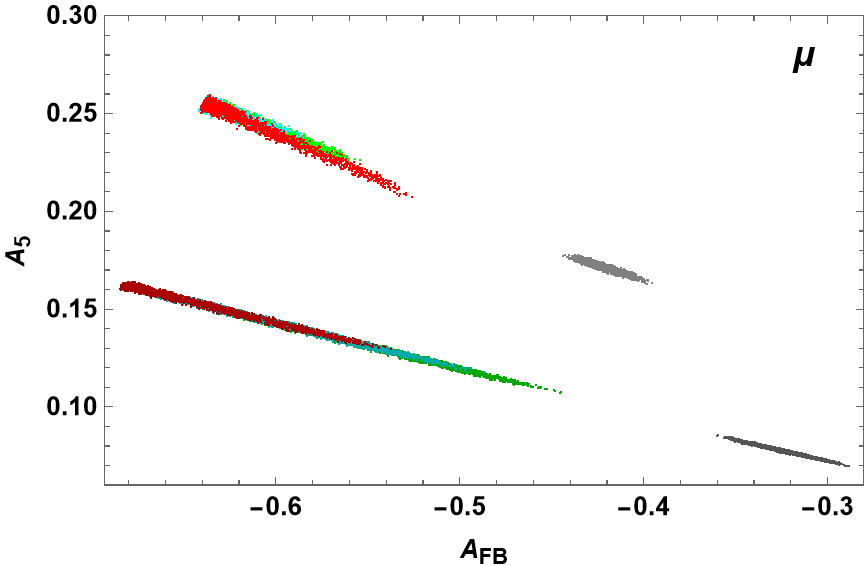}
 \includegraphics[width=2in]{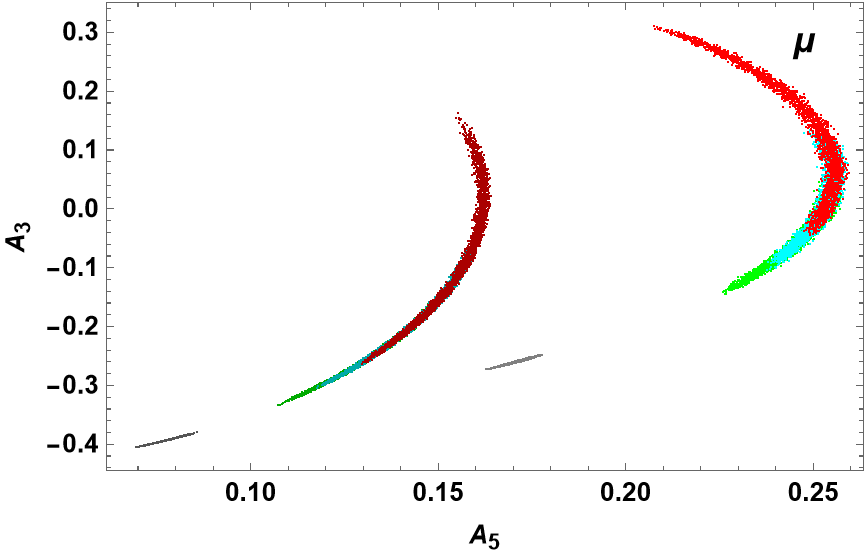}
 \includegraphics[width=2in]{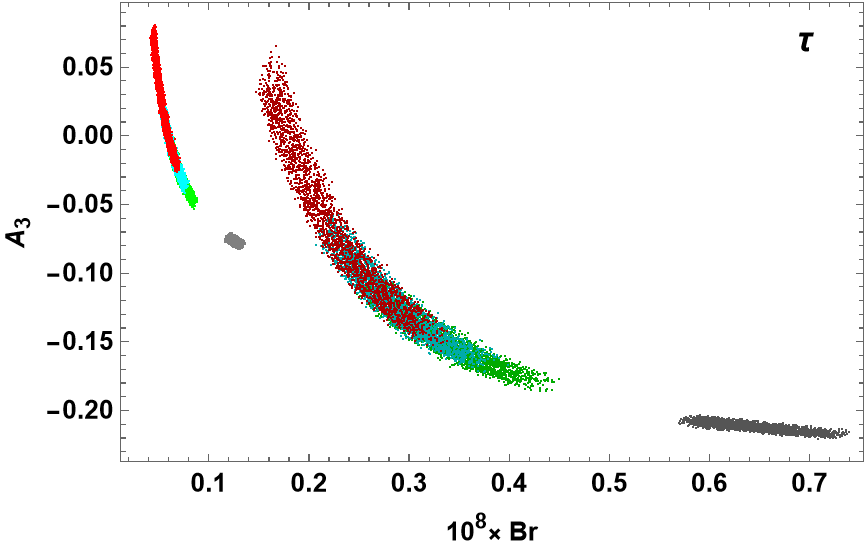}
 \includegraphics[width=2in]{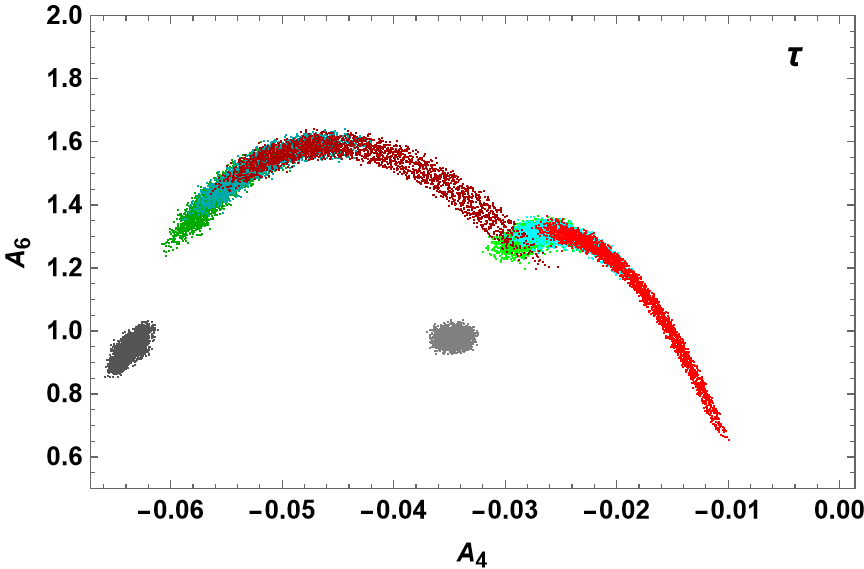}
 \includegraphics[width=2in]{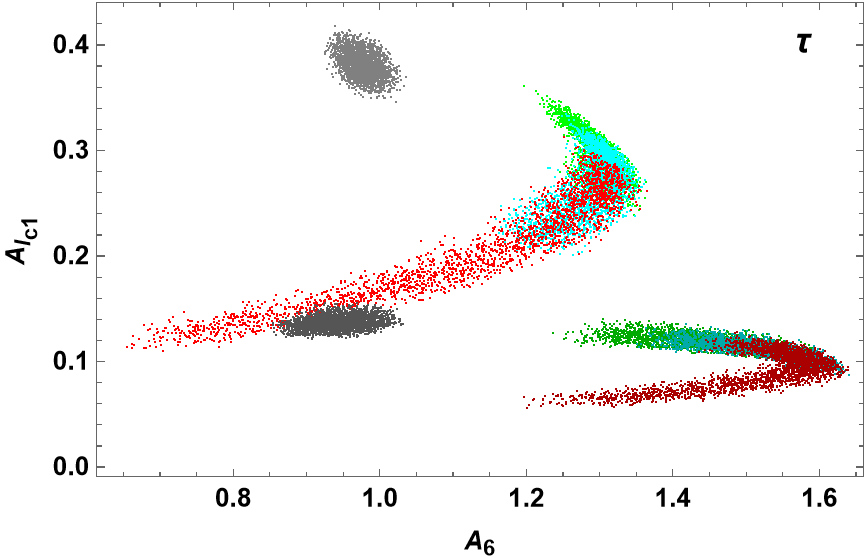}
 \caption{Correlations between different observables, for low bin (light color) and high bin (dark color) of the $B_s ^{*0} \rightarrow  D_s^-(\rightarrow \tau^-\,\bar\nu_{\tau})\ell^+ {\nu}_\ell$ decay.}
 \label{mu coralation}
\end{figure}
In order to be able to distinguish the benchmark points from each other, we have plotted correlations between the different physical observables. We show several correlations for the case of the $\mu$ ($\tau$) in the upper (lower) row of Figure~\ref{mu coralation}. In the case of the decay to the $\mu$, we show the correlations in the $(Br, A_4)$, $(A_{FB}, A_5)$ and $(A_5, A_3)$ planes. Along similar lines, for the case of the $\tau$, we show correlations in the $(Br, A_3)$, $(A_4, A_6)$ and $(A_6, A_{I_{c1}})$ planes. In these correlation plots, we have used the darker shade to represent the result in the high $q^2$ band while the lighter color represents the low $q^2$ result. We can see the SM occupy a distinct region in these plots in all planes, as expected. Further, we can see that some parameter space region inevitably overlaps among the benchmark points. However, we do see BVI separate from BII and BIV in several planes for most of its parameter space in both the low and high $q^2$ bins. This behavior is most pronounced in the case of the decay to the $\tau$ in the $(A_4, A_6)$ and $(A_6, A_{I_{c1})}$ planes.

\begin{table}[H]
\centering
\renewcommand{\arraystretch}{1.2}
\setlength{\tabcolsep}{8pt}
\resizebox{1\textwidth}{!}{
\begin{tabular}{|l|l|ccccccc|}\hline
& Models & $10^8\times \text{Br}$ & $-1\times A_{FB}$ & $A_{I_{c1}}$ & $A_3$ & $A_4$ & $A_5$ & $10^{-1} \times A_{6s}$ \\
\hline
& SM & $(0.36, 0.41)$ & $(0.38, 0.38)$ & $(0.30, 0.30)$ & $(-0.23, -0.23)$ & $(-0.16, -0.16)$ & $(0.18, 0.18)$ & $(0.17, 0.17)$ \\
\hline
\multirow{3}{*}{$C_V^L$} 
& EFT       & $(0.54, 0.82)$ & $--$ & $--$ & $--$ & $--$ & $--$ & $--$ \\
& LQ(4 TeV) & $(0.57, 0.88)$ & $--$ & $--$ & $--$ & $--$ & $--$ & $--$ \\
& LQ(2 TeV) & $(0.60, 0.98)$ & $--$ & $--$ & $--$ & $--$ & $--$ & $--$ \\
\hline
\multirow{3}{*}{$C_V^R$}
& EFT       & $(0.23, 0.25)$ & $(0.52, 0.58)$ & $(0.20, 0.26)$ & $(-0.11, 0.04)$ & $(-0.13,-0.10)$ & $(0.24, 0.27)$ & $(0.23, 0.26)$ \\
& LQ(4 TeV) & $(0.20, 0.23)$ & $(0.55, 0.58)$ & $(0.17, 0.24)$ & $(-0.07,0.12)$ & $(-0.12,-0.08)$ & $(0.25, 0.27)$ & $(0.24, 0.26)$ \\
& LQ(2 TeV) & $(0.18, 0.21)$ & $(0.45, 0.57)$ & $(0.10, 0.22)$ & $(-0.02, 0.30)$ & $(-0.11, -0.04)$ & $(0.24, 0.27)$ & $(0.25, 0.26)$ \\
\hline
\end{tabular}
}
\caption{Observables for the $B^{*0}_{s} \rightarrow 
\bigl[
   D_s^-(\rightarrow \tau^-\,\bar\nu_{\tau})
\bigr]\,
\mu^{+}\,{\nu}_\mu$ decay process in the low $q^2\in[s_{\text{min}},6]$GeV$^2$ region, comparing SM and NP models.}
\label{tab:bs_decay_mu_low}
\end{table}

\begin{table}[H]
\centering
\renewcommand{\arraystretch}{1.2}
\setlength{\tabcolsep}{8pt}
\resizebox{1\textwidth}{!}{
\begin{tabular}{|l|l|ccccccc|}
\hline 
& Models & $10^8\times \text{Br}$ & $-1\times A_{FB}$ & $A_{I_{c1}}$ & $A_3$ & $A_4$ & $A_5$ & $10^{-1} \times A_{6s}$ \\
\hline
& SM & $(0.41, 0.53)$ & $(0.29, 0.29)$ & $(0.08, 0.08)$ & $(-0.40, -0.40)$ & $(-0.12, -0.12)$ & $(0.06, 0.06)$ & $(0.13, 0.13)$ \\
\hline
\multirow{3}{*}{$C_V^L$} 
& EFT       & $(0.62, 1.06)$ & $--$ & $--$ & $--$ & $--$ & $--$ & $--$ \\
& LQ(4 TeV) & $(0.65, 1.13)$ & $--$ & $--$ & $--$ & $--$ & $--$ & $--$ \\
& LQ(2 TeV) & $(0.69, 1.26)$ & $--$ & $--$ & $--$ & $--$ & $--$ & $--$ \\
\hline
\multirow{3}{*}{$C_V^R$}
& EFT       & $(0.23, 0.26)$ & $(0.44, 0.56)$ & $(0.07, 0.08)$ & $(-0.33, -0.23)$ & $(-0.11, 0.01)$ & $(0.10, 0.13)$ & $(0.19, 0.25)$ \\
& LQ(4 TeV) & $(0.19, 0.23)$ & $(0.48, 0.61)$ & $(0.06, 0.08)$ & $(-0.31, -0.16)$ & $(-0.11,-0.08)$ & $(0.11, 0.14)$ & $(0.21, 0.27)$ \\
& LQ(2 TeV) & $(0.14, 0.20)$ & $(0.52, 0.65)$ & $(0.04, 0.07)$ & $(-0.27, 0.03)$ & $(-0.10, -0.07)$ & $(0.12, 0.15)$ & $(0.23, 0.28)$ \\
\hline
\end{tabular}
}
\caption{Observables for the $B^{*0}_{s} \rightarrow 
\bigl[
   D_s^-(\rightarrow \tau^-\,\bar\nu_{\tau})
\bigr]\,
\mu^{+}\,{\nu}_\mu$ decay process in the high $q^2\in[6,s_{\text{max}}]$GeV$^2$ region, comparing SM and NP models.}
\label{tab:bs_decay_mu_high}
\end{table}

\begin{table}[H]
\centering
\renewcommand{\arraystretch}{1.2}
\setlength{\tabcolsep}{8pt}
\resizebox{1\textwidth}{!}{
\begin{tabular}{|l|l|ccccccc|}
\hline 
& Models & $10^8\times \text{Br}$ & $-1\times A_{FB}$ & $A_{I_{c1}}$ & $A_3$ & $A_4$ & $A_5$ & $10^{-1} \times A_{6s}$ \\
\hline
& SM & $(0.04, 0.04)$ & $(0.08, 0.08)$ & $(0.45, 0.45)$ & $(-0.05, -0.05)$ & $(-0.02, -0.02)$ & $(0.33, 0.33)$ & $(0.11, 0.11)$ \\
\hline
\multirow{3}{*}{$C_V^L$}
& EFT       & $(0.05, 0.09)$ & $--$ & $--$ & $--$ & $--$ & $--$ & $--$ \\
& LQ(4 TeV) & $(0.06, 0.09)$ & $--$ & $--$ & $--$ & $--$ & $--$ & $--$ \\
& LQ(2 TeV) & $(0.06, 0.10)$ & $--$ & $--$ & $--$ & $--$ & $--$ & $--$ \\
\hline
\multirow{3}{*}{$C_V^R$}
& EFT       & $(0.02, 0.02)$ & $(0.21, 0.31)$ & $(0.31, 0.40)$ & $(-0.00, -0.03)$ & $(-0.02, -0.01)$ & $(0.30, 0.33)$ & $(0.16, 0.19)$ \\
& LQ(4 TeV) & $(0.02, 0.02)$ & $(0.24, 0.34)$ & $(0.26, 0.37)$ & $(-0.02, 0.02)$ & $(-0.02, -0.01)$ & $(0.28, 0.33)$ & $(0.17, 0.19)$ \\
& LQ(2 TeV) & $(0.01, 0.02)$ & $(0.28, 0.34)$ & $(0.14, 0.34)$ & $(-0.01, 0.05)$ & $(-0.02, -0.01)$ & $(0.20, 0.32)$ & $(0.17, 0.18)$ \\
\hline
\end{tabular}
}
\caption{Observables for the $B^{*0}_{s} \rightarrow 
\bigl[
   D_s^-(\rightarrow \tau^-\,\bar\nu_{\tau})
\bigr]\,
\tau^{+}\,{\nu}_\tau$ decay process in the low $q^2\in[s_{\text{min}},6]$GeV$^2$ region, comparing SM and NP models.}
\label{tab:bs_decay_tau_low}
\end{table}

\begin{table}[H]
\centering
\renewcommand{\arraystretch}{1.2}
\setlength{\tabcolsep}{8pt}
\resizebox{1\textwidth}{!}{
\begin{tabular}{|l|l|ccccccc|}
\hline 
& Models & $10^8\times \text{Br}$ & $-1\times A_{FB}$ & $A_{I_{c1}}$ & $A_3$ & $A_4$ & $A_5$ & $10^{-1} \times A_{6s}$ \\
\hline
& SM & $(0.19, 0.25)$ & $(0.19, 0.19)$ & $(0.13, 0.13)$ & $(-0.21, -0.21)$ & $(-0.06, -0.06)$ & $(0.15, 0.15)$ & $(0.10, 0.10)$ \\
\hline
\multirow{3}{*}{$C_V^L$}
& EFT       & $(0.29, 0.50)$ & $--$ & $--$ & $--$ & $--$ & $--$ & $--$ \\
& LQ(4 TeV) & $(0.31, 0.54)$ & $--$ & $--$ & $--$ & $--$ & $--$ & $--$ \\
& LQ(2 TeV) & $(0.33, 0.60)$ & $--$ & $--$ & $--$ & $--$ & $--$ & $--$ \\
\hline
\multirow{3}{*}{$C_V^R$}
& EFT       & $(0.10, 0.12)$ & $(0.31, 0.42)$ & $(0.10, 0.12)$ & $(-0.18, -0.13)$ & $(-0.06, -0.05)$ & $(0.17, 0.18)$ & $(0.16, 0.20)$ \\
& LQ(4 TeV) & $(0.09, 0.11)$ & $(0.47, 0.48)$ & $(0.09, 0.11)$ & $(-0.17, -0.10)$ & $(-0.06,- 0.04)$ & $(0.17, 0.18)$ & $(0.17, 0.22)$ \\
& LQ(2 TeV) & $(0.06, 0.09)$ & $(0.38, 0.52)$ & $(0.05, 0.10)$ & $(-0.15, 0.02)$ & $(-0.05, -0.03)$ & $(0.16, 0.18)$ & $(0.19, 0.24)$ \\
\hline
\end{tabular}
}
\caption{Observables for the $B^{*0}_{s} \rightarrow 
\bigl[
   D_s^-(\rightarrow \tau^-\,\bar\nu_{\tau})
\bigr]\,
\tau^{+}\,{\nu}_\tau$ decay process in the high $q^2\in[6,s_{\text{max}}]$GeV$^2$ region, comparing SM and NP models.}
\label{tab:bs_decay_tau_high}
\end{table}

\subsubsection{Semi-analytic expressions for the observables}

We now provide semi-analytic expressions for the various physical observables. 
The expression for the general observable $\mathcal{O}_i$ is given by,
\begin{align}
 \mathcal{O}_i&=\mathcal{B}_1
  +  \mathcal{B}_2 C_V^L
  +  \mathcal{B}_3 \bigl(C_V^L\bigr)^2
 + \mathcal{B}_4 C_V^R
  +  \mathcal{B}_5 \bigl(C_V^R\bigr)^2+ \mathcal{B}_6 C_V^L \, C_V^R,
\label{semi-exp}\end{align}
where the $\mathcal{B}_i$'s are coefficients that are given in Table~\ref{tab:lowmuon-observables} (low $q^2$ bin, decay to $\mu$), Table~\ref{tab:highmuon-observables} (high $q^2$ bin, decay to $\mu$), Table~\ref{tab:lowtau-observables} (low $q^2$ bin, decay to $\tau$) and Table~\ref{hightau-observables} (high $q^2$ bin, decay to $\tau$). 
\begin{table}[H]
\centering
\begin{tabular}{|c|c|c|c|c|c|c|}
\hline \hline
\textbf{$\mathcal{O}_i$} & $\mathcal{B}_1$ & $\mathcal{B}_2$ & $\mathcal{B}_3$ & $\mathcal{B}_4$ & $\mathcal{B}_5$ & $\mathcal{B}_6$ \\
\hline
$\text{Br}\times 10^{8}$ & $0.38 $ & $0.77$ & $0.38$ & $-0.58$ & $0.38$ & $-0.58$ \\
$A_{FB}$ & $-0.37$ & $0$ & $0$ & $-0.59$ & $-0.78$ & $-0.59$ \\
$A_4$ & $-0.16$ & $0$ & $0$ & $0.077 $ & $0$ & $0.077$ \\
$A_5$ & $0.18$ & $0$ & $0$ & $0.26$ & $-0.34$ & $0.26$ \\
$A_3$ & $-0.23$ & $0$ & $0$ & $0.35$ & $0$ & $0.35$ \\
$A_{6s}$ & $1.73$ & $0$ & $0$ & $2.66$ & $-3.54$ & $2.66$ \\
$A_{I_{c1}}$ & $0.30 $ & $0$ & $0$ & $-0.12 $ & $0$ & $-0.12$ \\
\hline
\end{tabular}
\caption{Values for the constants $\mathcal{B}_i$ for $B^{*0}_{s}\to\bigl[D_s^-(\rightarrow \tau^-\,\bar\nu_{\tau})\bigr]\,\mu^{+}\,{\nu}_\mu$ observables in low $q^2\in (s_{min}, 6~{\rm GeV^2})$ bin, used in the expression~\ref{semi-exp}.}
\label{tab:lowmuon-observables}
\end{table}

\begin{table}[H]
\centering
\begin{tabular}{|c|c|c|c|c|c|c|}
\hline \hline
\textbf{$\mathcal{O}_i$} & $\mathcal{B}_1$ & $\mathcal{B}_2$ & $\mathcal{B}_3$ & $\mathcal{B}_4$ & $\mathcal{B}_5$ & $\mathcal{B}_6$ \\
\hline
$\text{Br}\times 10^8$ & $0.46$ & $0.93$ & $0.46$ & $-0.82$ & $0.46$ & $-0.82$ \\
$A_{FB}$ & $-0.29$ & $0$ & $0$ &$ -0.51$ & $-0.59 $ & $-0.51$ \\
$A_4$ & $-0.12 $ & $0$ & $0$ & $0.03 $ & $0$ & $0.03$ \\
$A_5$ & $0.06$ & $0$ & $0$ & $0.12$ & $-0.13$ & $0.12$ \\
$A_3$ & $-0.40 $ & $0$ & $0$ & $0.18$ & $0$ & $0.18$ \\
$A_{6s}$ & $1.33$ & $0$ & $0$ & $2.33$ & $-2.66$ & $2.33$ \\
$A_{I_{c1}}$ & $0.08$ & $0$ & $0$ & $-0.02$ & $0$ & $-0.02$ \\
\hline
\end{tabular}
\caption{Values for the constants $\mathcal{B}_i$ for $B^{*0}_{s} \rightarrow \bigl[ D_s^-(\rightarrow \tau^-\,\bar\nu_{\tau})\bigr]\,\mu^{+}\,{\nu}_\mu$ observables in high $q^2\in (6~{\rm GeV^2} ,s_{max})$ bin, used in the expression~\ref{semi-exp}. }
\label{tab:highmuon-observables}
\end{table}

\begin{table}[H]
\centering
\begin{tabular}{|c|c|c|c|c|c|c|}
\hline \hline
\textbf{$\mathcal{O}_i$} &$ \mathcal{B}_1 $ &  $ \mathcal{B}_2 $& $ \mathcal{B}_3 $ & $ \mathcal{B}_4$ &  $ \mathcal{B}_5 $ &  $ \mathcal{B}_6 $ \\
\hline 
$\text{Br}\times 10^8$&$0.04$&$0.08$&$0.04$&$-0.06$&$ -0.04$&$-0.06$ \\$A_{FB}$&$-0.08$&$0$&$0$&$-0.17$&$ -0.20$&$-0.17$ \\
$A_4$&$-0.02$&$0$&$0$&$0.01$&$0$&$0.01$ \\
$A_5$&$0.33$&$0$&$0$&$0.04$&$-0.16$&$0.04$  \\
$A_3$&$-0.05$&$0$&$0$&$0.06$&$0$& $0.06$ \\
$A_{6s}$&$1.15$&$0$&$0$&$1.87$&$-2.31$& $1.87$\\
$A_{I_{c1}}$&$0.46$&$0$&$0$&$-0.17$&$0$& $-0.17$\\
\hline
\end{tabular}
\caption{Values for the constants $\mathcal{B}_i$ for $B^{*0}_{s}\to\bigl[D_s^-(\rightarrow \tau^-\,\bar\nu_{\tau})\bigr]\,\tau^{+}\,{\nu}_\tau$ observables in low $q^2\in (s_{min}, 6~{\rm GeV^2})$ bin, used in the expression~\ref{semi-exp}. }
\label{tab:lowtau-observables}
\end{table}

\begin{table}[H]
\centering
\begin{tabular}{|c|c|c|c|c|c|c|}
\hline \hline
\textbf{$\mathcal{O}_i$} &$ \mathcal{B}_1 $ &  $ \mathcal{B}_2 $& $ \mathcal{B}_3 $ & $ \mathcal{B}_4$ &  $ \mathcal{B}_5 $ &  $ \mathcal{B}_6 $ \\
\hline 
$\text{Br}\times 10^8$&$0.22$&$0.44$&$0.22$&$-0.39$&$0.22$&$-0.39$\\$
A_{FB}$&$-0.19$&$0$&$0$&$-0.37$&$-0.41$&$-0.37$ \\$A_4$&$-0.06$&$0$&$0$&$0.01$&$0$&$0.01$ \\
$A_5$&$0.15$&$0$&$0$&$0.07$&$-0.11$&$0.07$\\
$A_3$&$-0.21$&$0$&$0$&$0.08$&$0$&$0.08$\\
$A_{6s}$&$1.07$&$0$&$0$&$1.90$&$-2.15$&$1.90$\\
$A_{I_{c1}}$&$0.13$&$0$&$0$&$0.03$&$0$&$0.03$\\
\hline
\end{tabular}
\caption{Values for the constants $\mathcal{B}_i$ for $B^{*0}_{s} \rightarrow \bigl[ D_s^-(\rightarrow \tau^-\,\bar\nu_{\tau})\bigr]\,\tau^{+}\,{\nu}_\tau$ observables in high $q^2\in (6~{\rm GeV^2} ,s_{max})$ bin, used in the expression~\ref{semi-exp}. }
\label{hightau-observables}
\end{table}

\subsection{Chirality structure of charged current vector-like NP in the $c\to s$ transition}\label{cTos_transition}
\begin{figure}[H]
\centering
\includegraphics[width=1.5in]{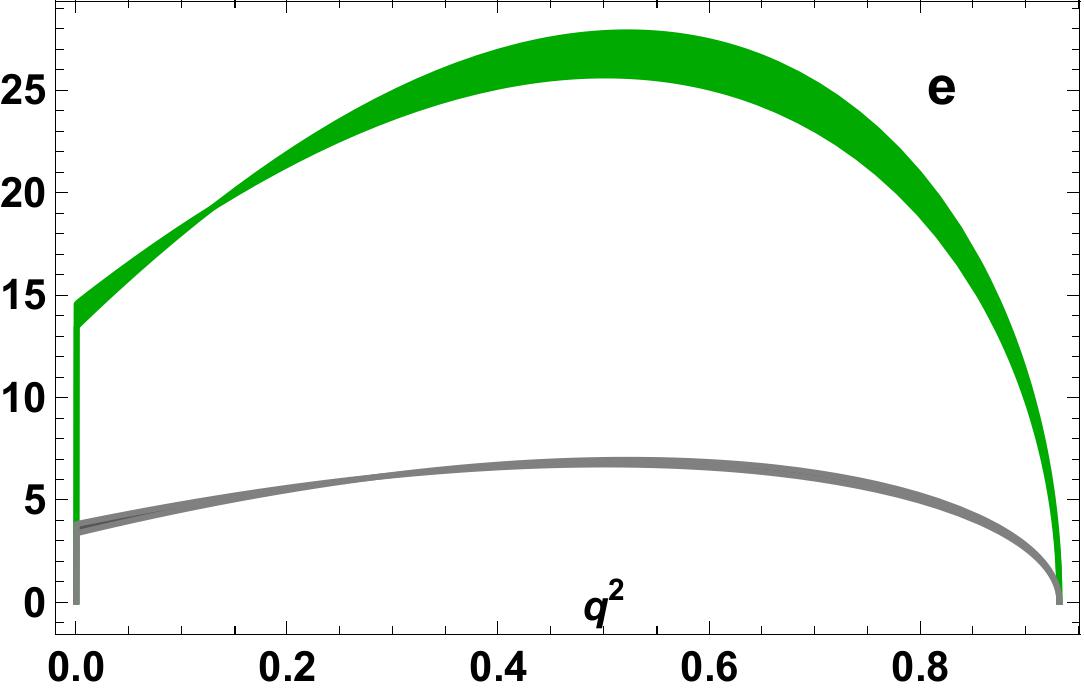} 
\includegraphics[width=1.5in]{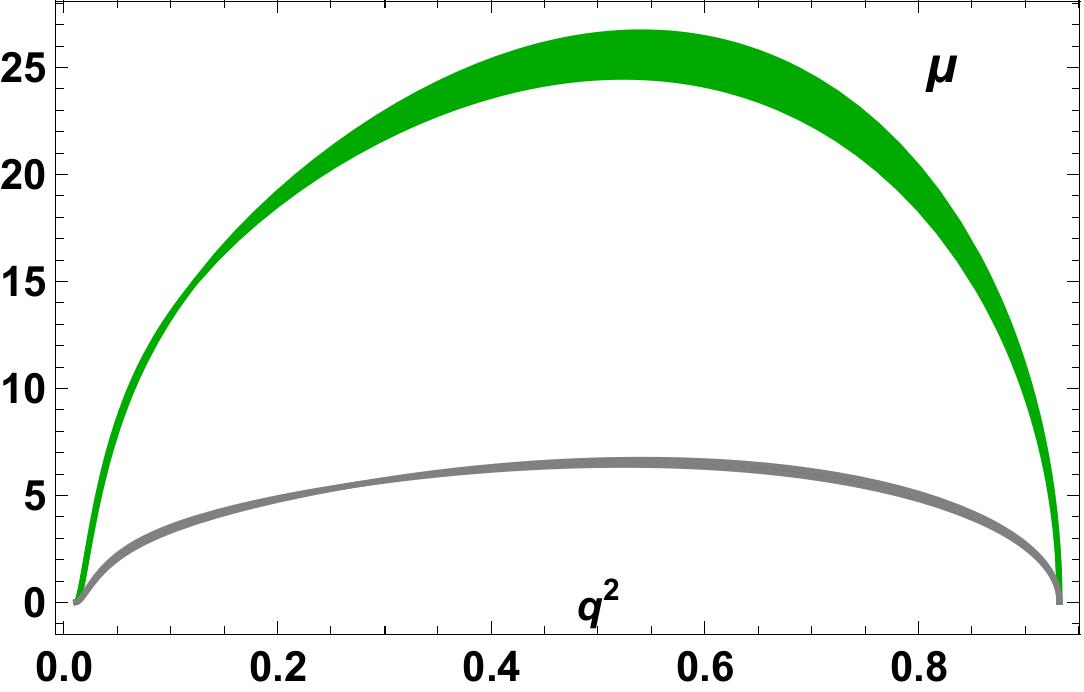}
\includegraphics[width=1.5in]{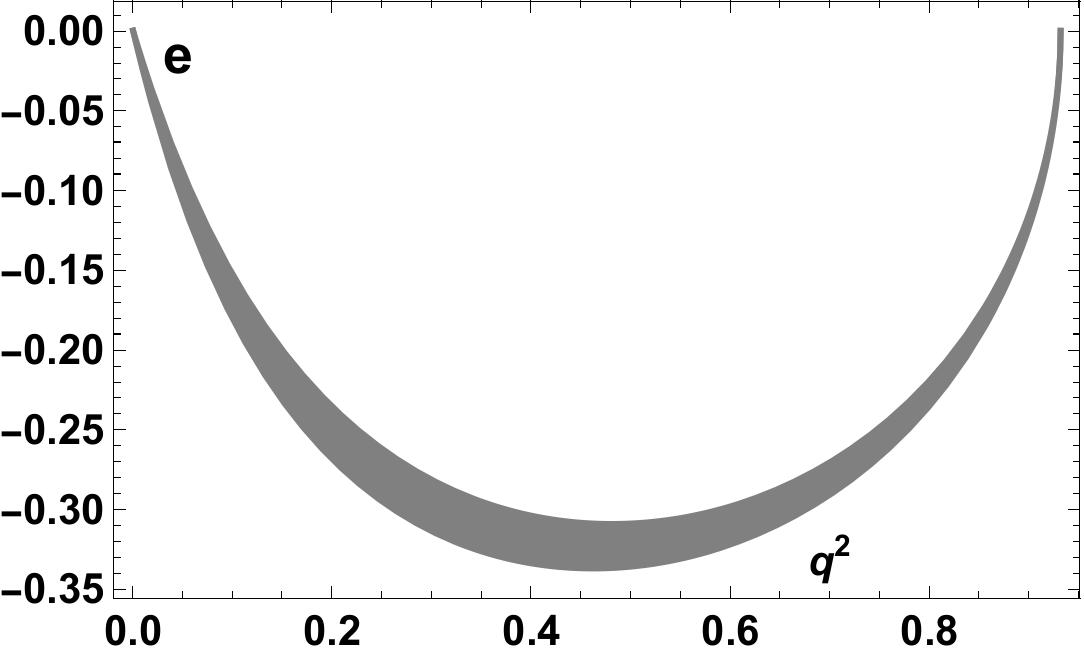}
\includegraphics[width=1.5in]{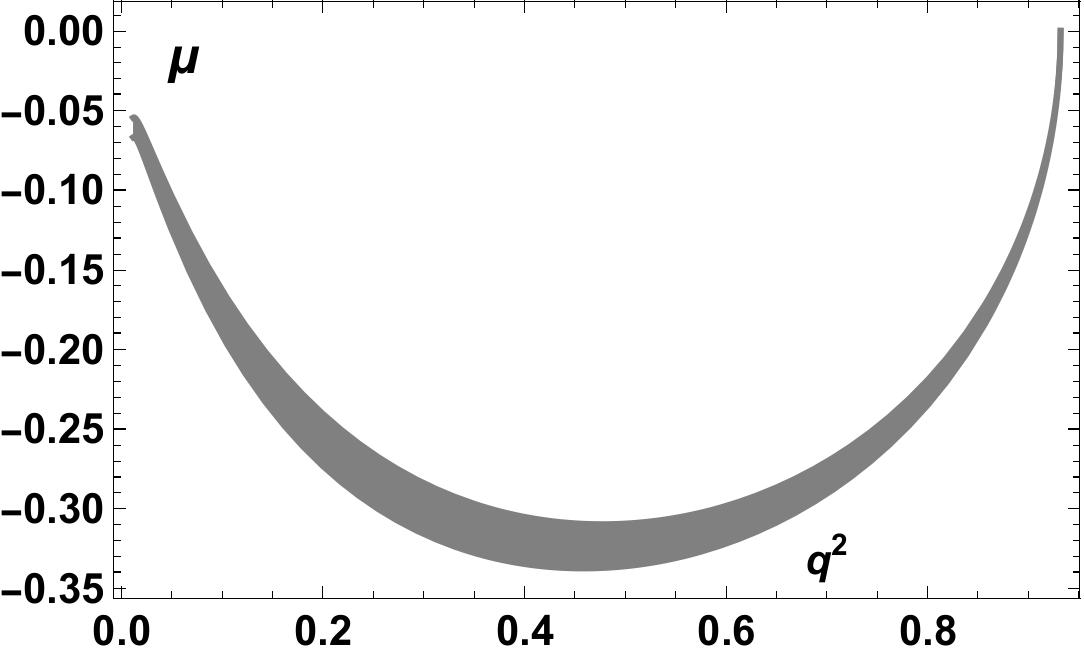}
\caption{$\text{Br}\times 10^8$ (first two plots) and $A_{FB}$ (last two plots) of $B_c^{*+} \rightarrow B_s^0(\to D_s^{*-} \mu^+\nu_\mu) \ell^{+}{\nu}_\ell$.}
\label{BrVsq2 BsDs}
\end{figure}

We have noted that for the $b\to c$ transition, the angular observables are not sensitive to the left-handed charged current contribution, $C_{V_L}$. We expect a similar result for other charged current transitions, \emph{i.e.} we expect an NP contribution to the Br 
from $C_{V_L}$ with no corresponding contribution to either the $A_{FB}$ or the various angular observables 
that we have computed. In particular, we have computed the case for the 
$c\to s$ transition in the decays $B^{*+}_c \rightarrow 
\left[ P(\rightarrow P'\,\mu^+\,\nu_{\mu}) 
\right]\,
\ell^{+}\,{\nu}_\ell$ where $P$ is $B_s^0$ ($D^0$) and $P'$ is $D_s^{*-}$ ($K^-$). For the purposes of this computation, we have 
taken the scenario $C_{V_L}=[0.957,\,1.002]$~\cite{Iguro:2024hyk}. 
{We used the same theoretical description for the $c\to s$ transition decays by incorporating the form factor extrapolation mentioned in Eq~\ref{ff eq}, and its values are given in Table~\ref{ff table 2} with the initial mass $m=m_{B^*_c }=6.332$. }
In Figure~\ref{BrVsq2 BsDs} we show the result for the branching ratio and the 
forward-backward asymmetry for the decay $B^{*+}_c \rightarrow 
\left[ B_s^0(\rightarrow D_s^{*-}\,\mu^+\,\nu_{\mu}) 
\right]\,
\ell^{+}\,{\nu}_\ell$ in cases with $e$ and $\mu$ final states. The SM result is shown in gray with the NP contribution shown in green. We can see that as expected 
the branching ratio does receive an NP contribution from $C_{V_L}$ but no such contribution exists for the forward-backward asymmetry. Consequently, we only show the SM result for this. 

\begin{figure}[H]
\centering
\includegraphics[width=1.5in]{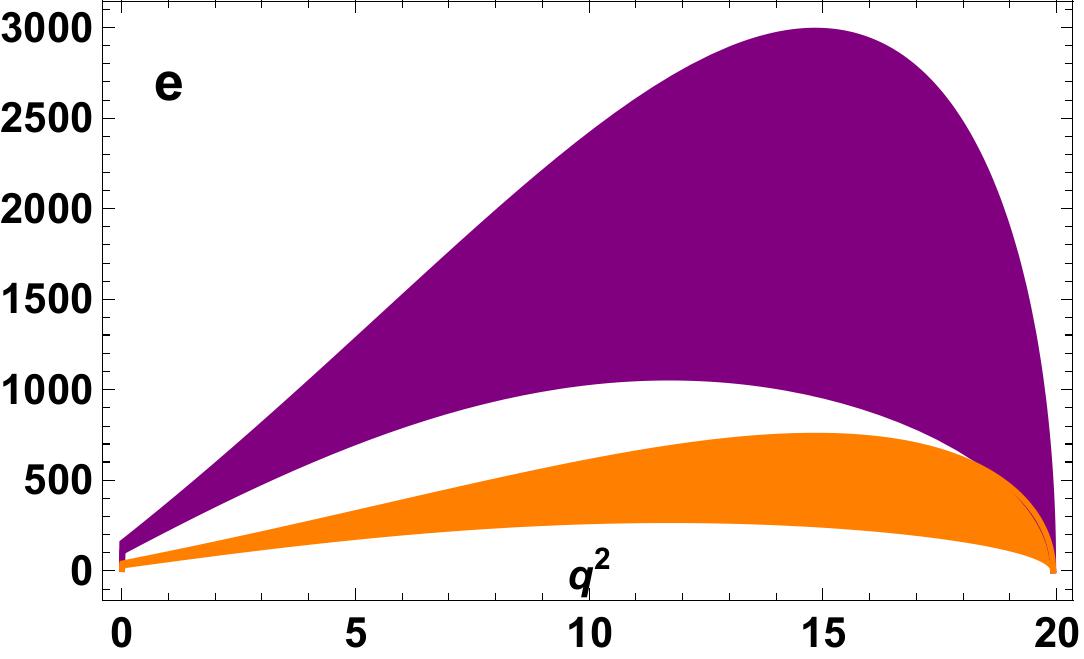} 
\includegraphics[width=1.5in]{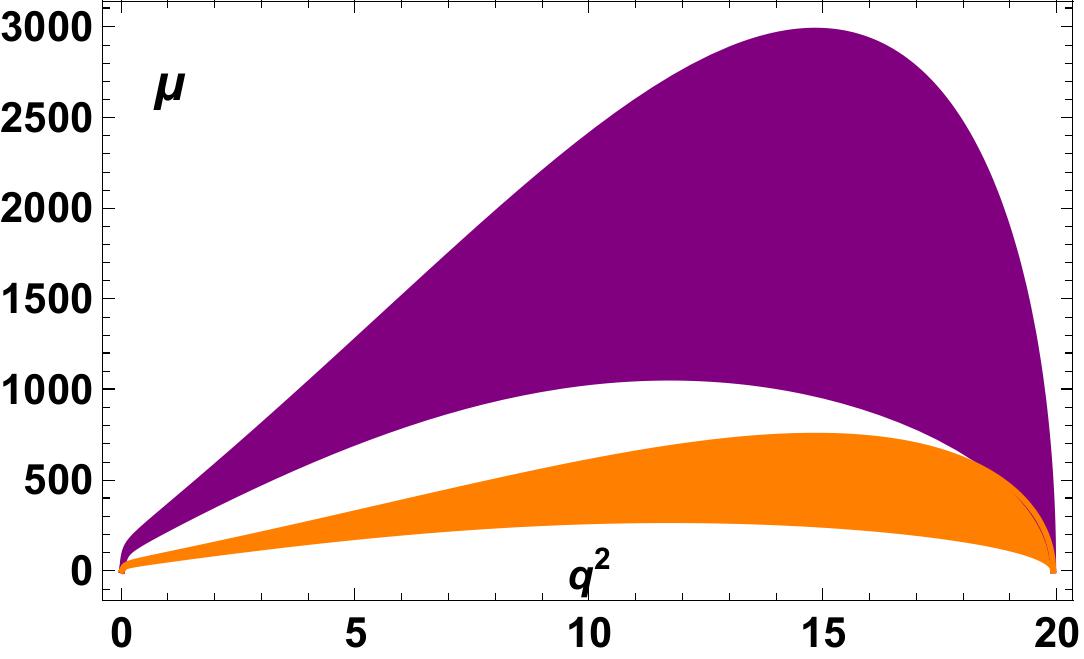}
\includegraphics[width=1.5in]{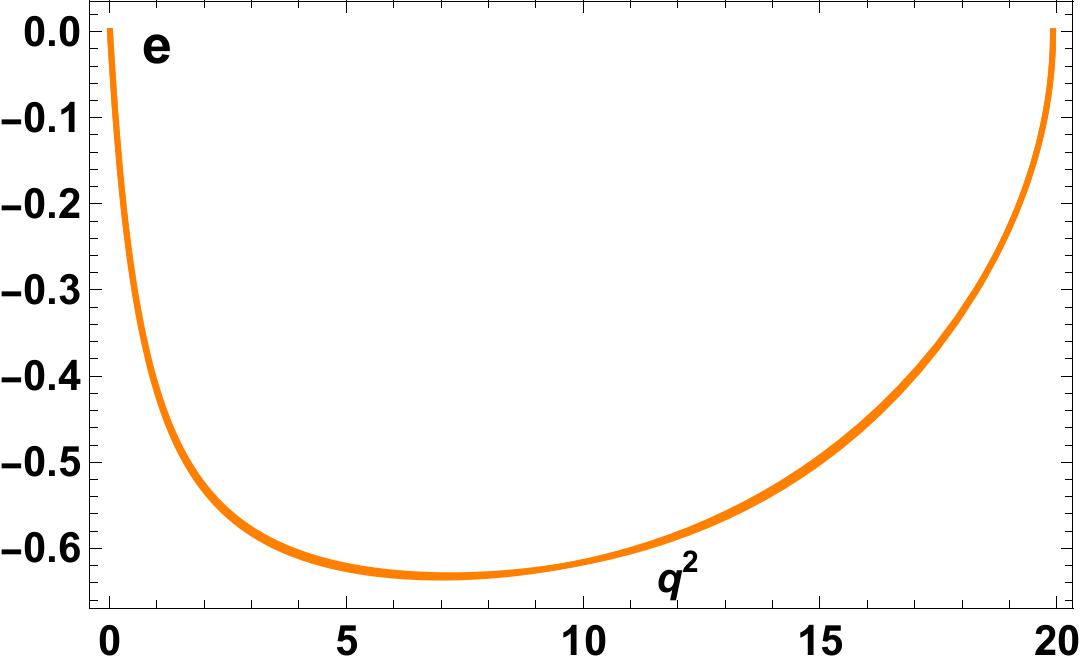}
\includegraphics[width=1.5in]{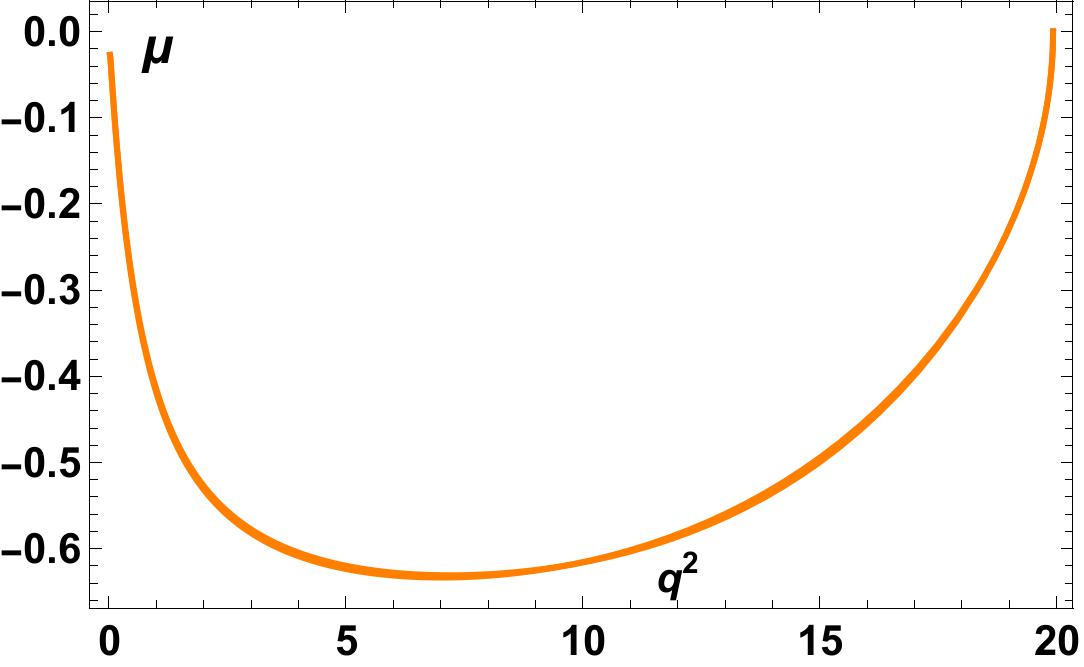}
\caption{$\text{Br}\times 10^8$ (first two plots) and $A_{FB}$ (last two plots) of $B_c^{*+} \rightarrow D^0(\to K^{-} \mu^+\nu_\mu) \ell^{+} {\nu}_\ell$.}
\label{BrVsq2 BcD0}
\end{figure}

A similar trend is seen in Figure~\ref{BrVsq2 BcD0} where we plot the result for the $B^{*+}_c \rightarrow 
\left[ D^0(\rightarrow K^-\,\mu^+\,\nu_{\mu}) 
\right]\,
\ell^{+}\,{\nu}_\ell$ decay. We again show only the SM result (orange) with the NP contribution shown in purple, and in the case of the forward-backward asymmetry, as again 
the $C_{V_L}$ contribution is conspicuous by its absence.

\begin{figure}[H]
\centering
\includegraphics[width=1.5in,height=1.5in]{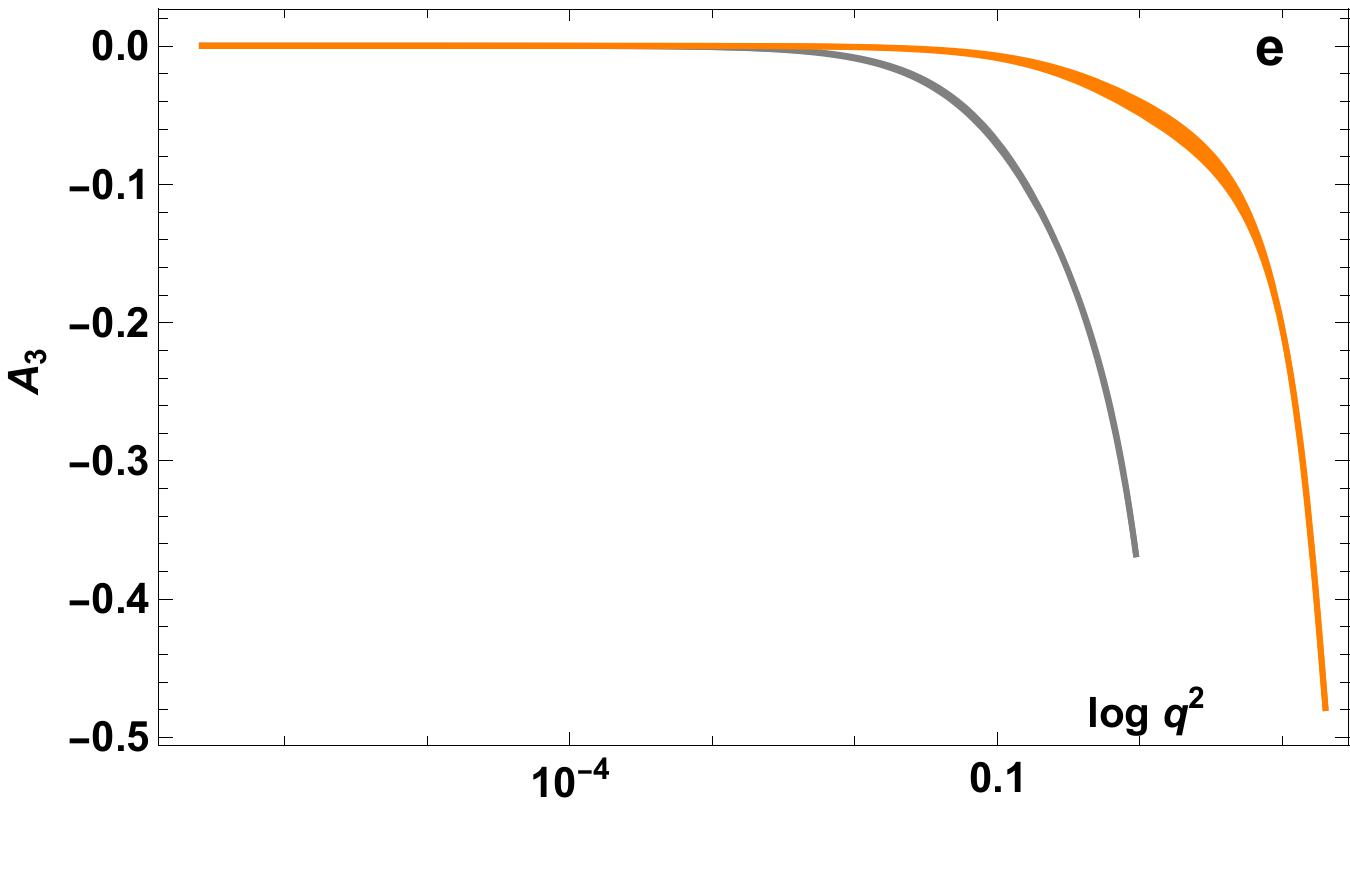}
\includegraphics[width=1.5in,height=1.5in]{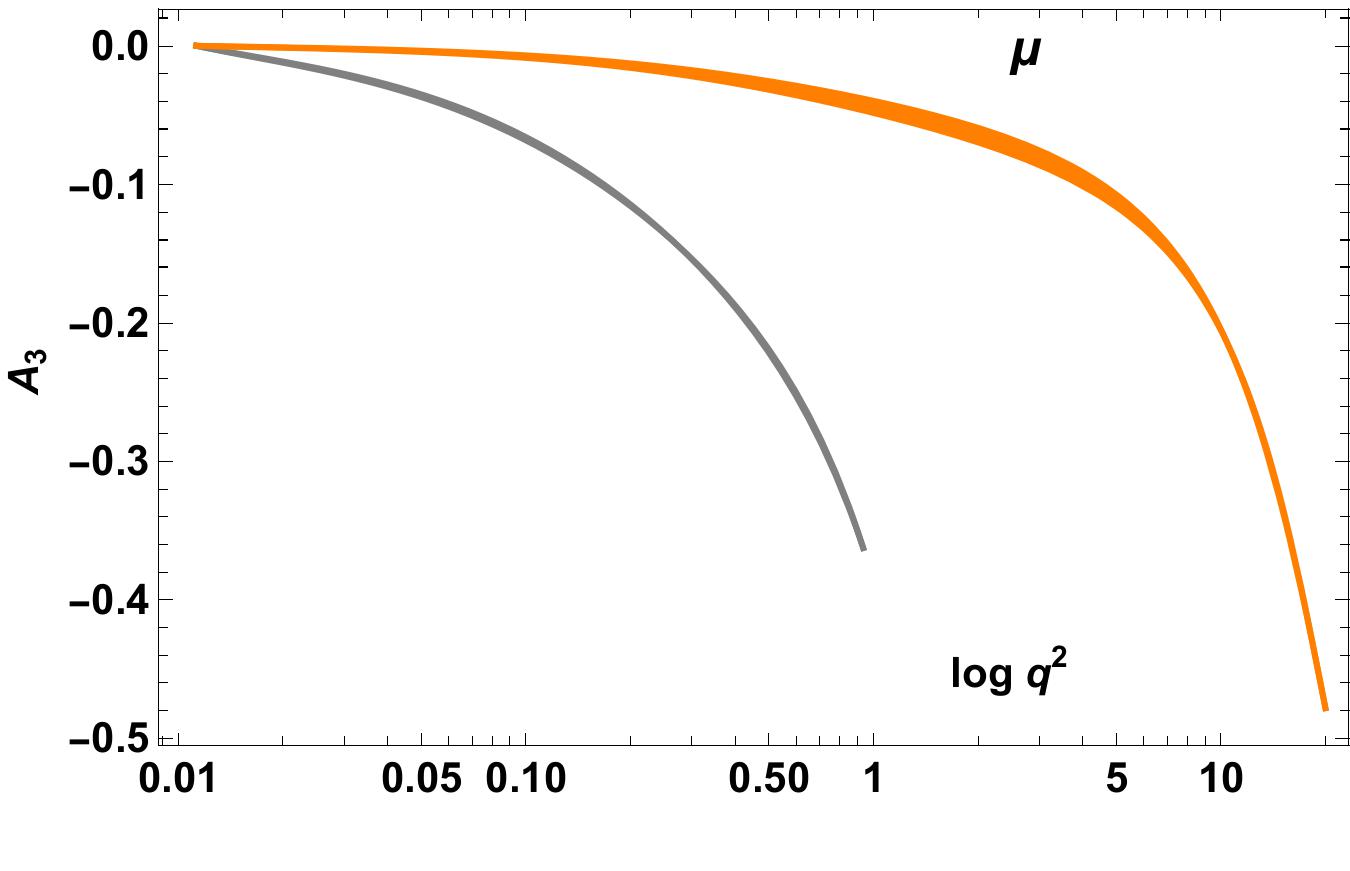}
\includegraphics[width=1.5in,height=1.5in]{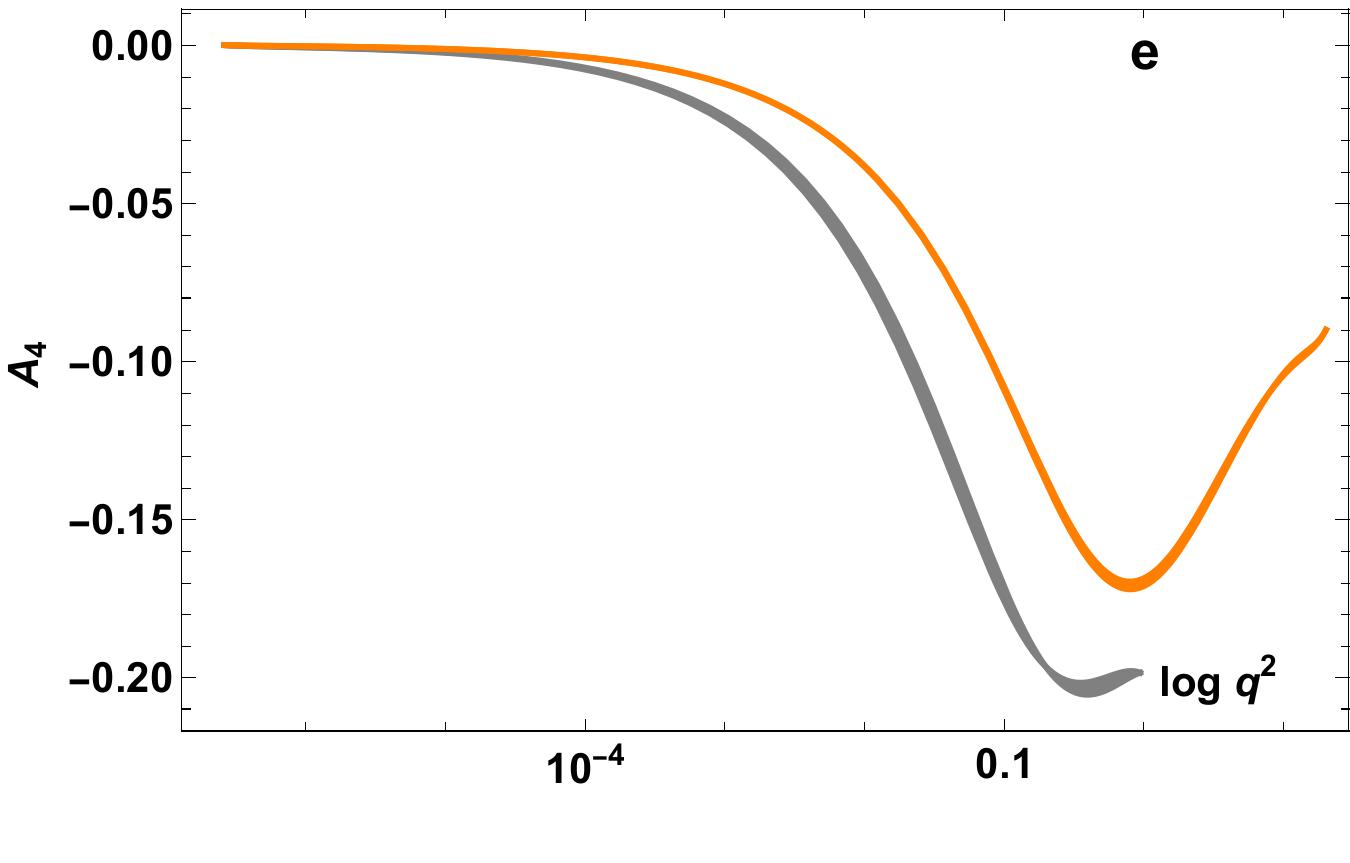}
\includegraphics[width=1.5in,height=1.5in]{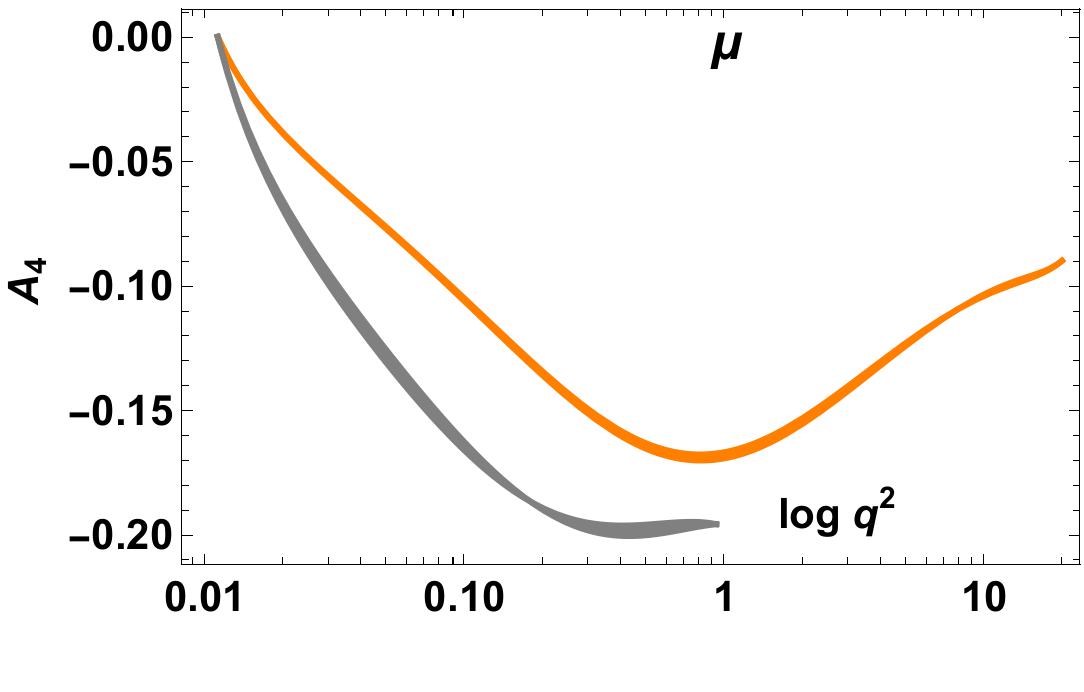}
\includegraphics[width=1.5in,height=1.5in]{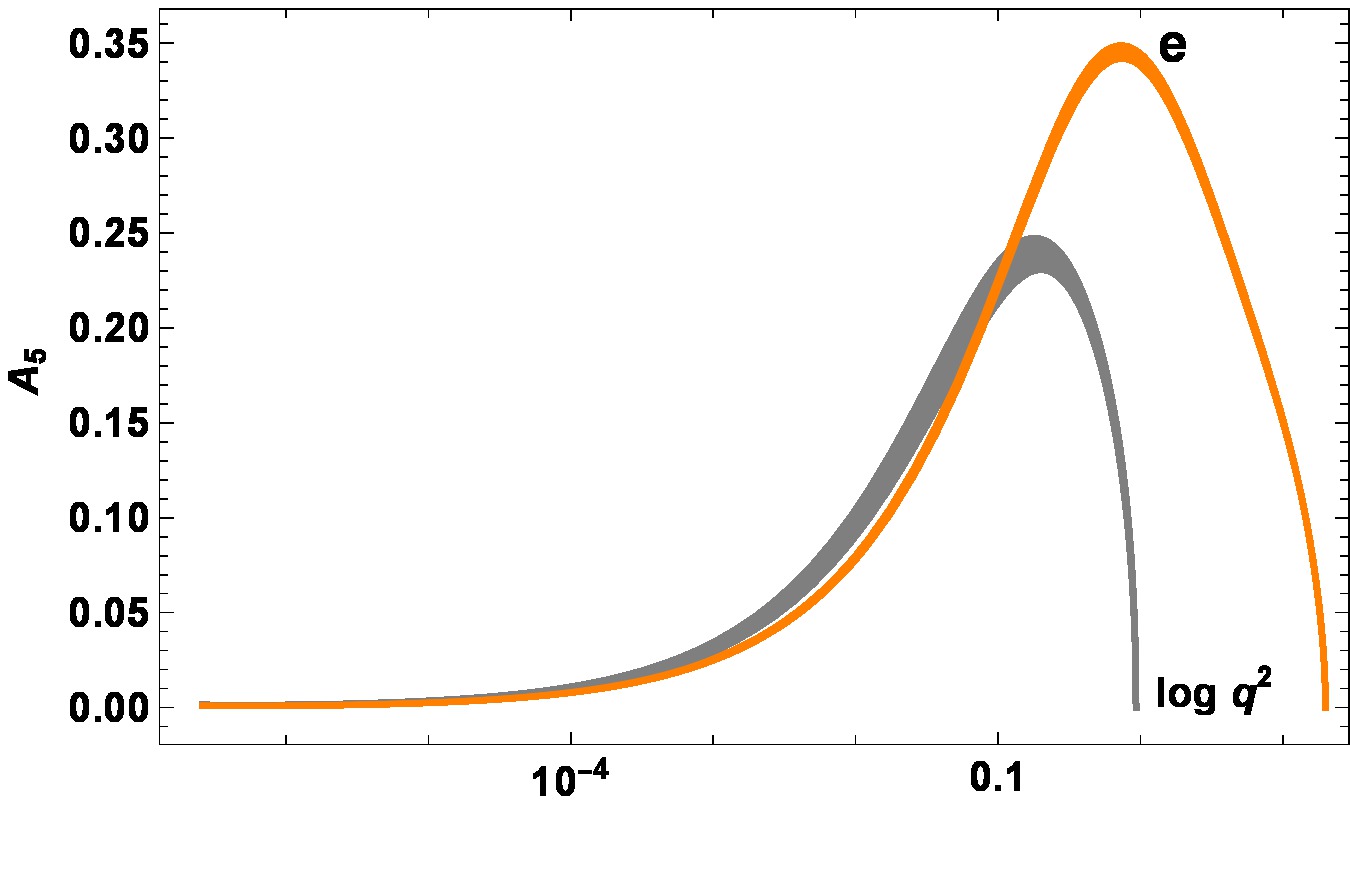}
\includegraphics[width=1.5in,height=1.5in]{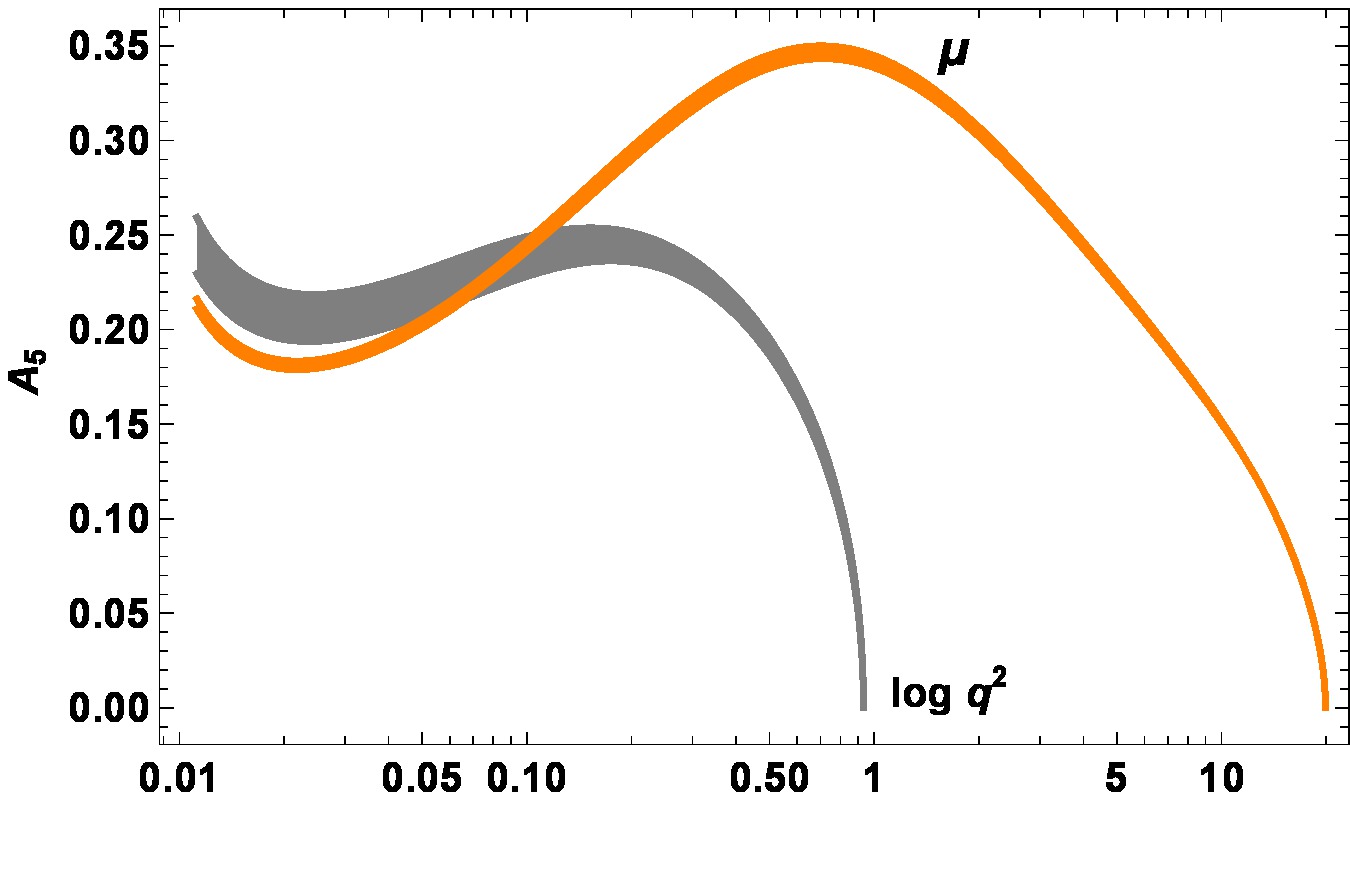}
\includegraphics[width=1.5in,height=1.5in]{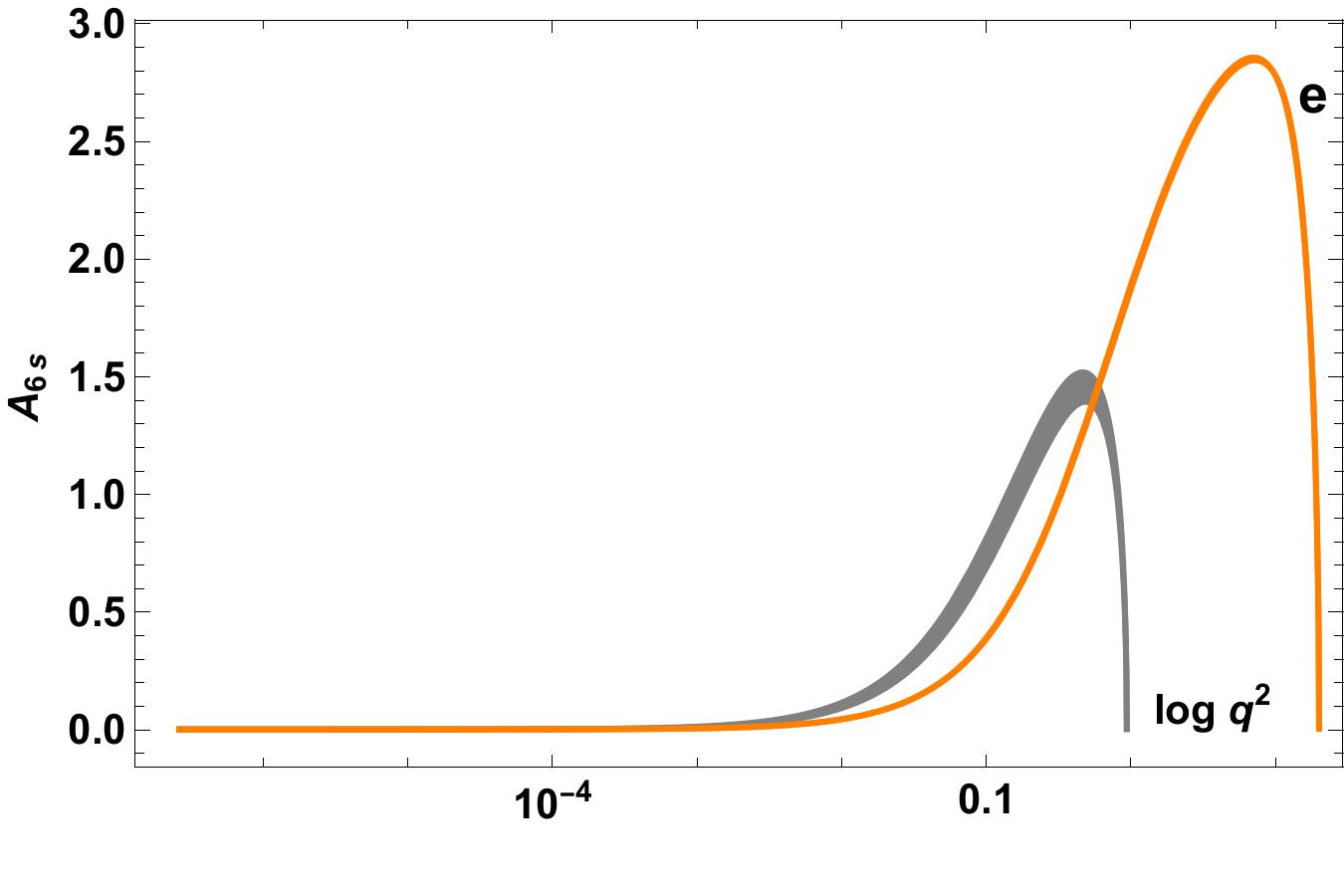}
\includegraphics[width=1.5in,height=1.5in]{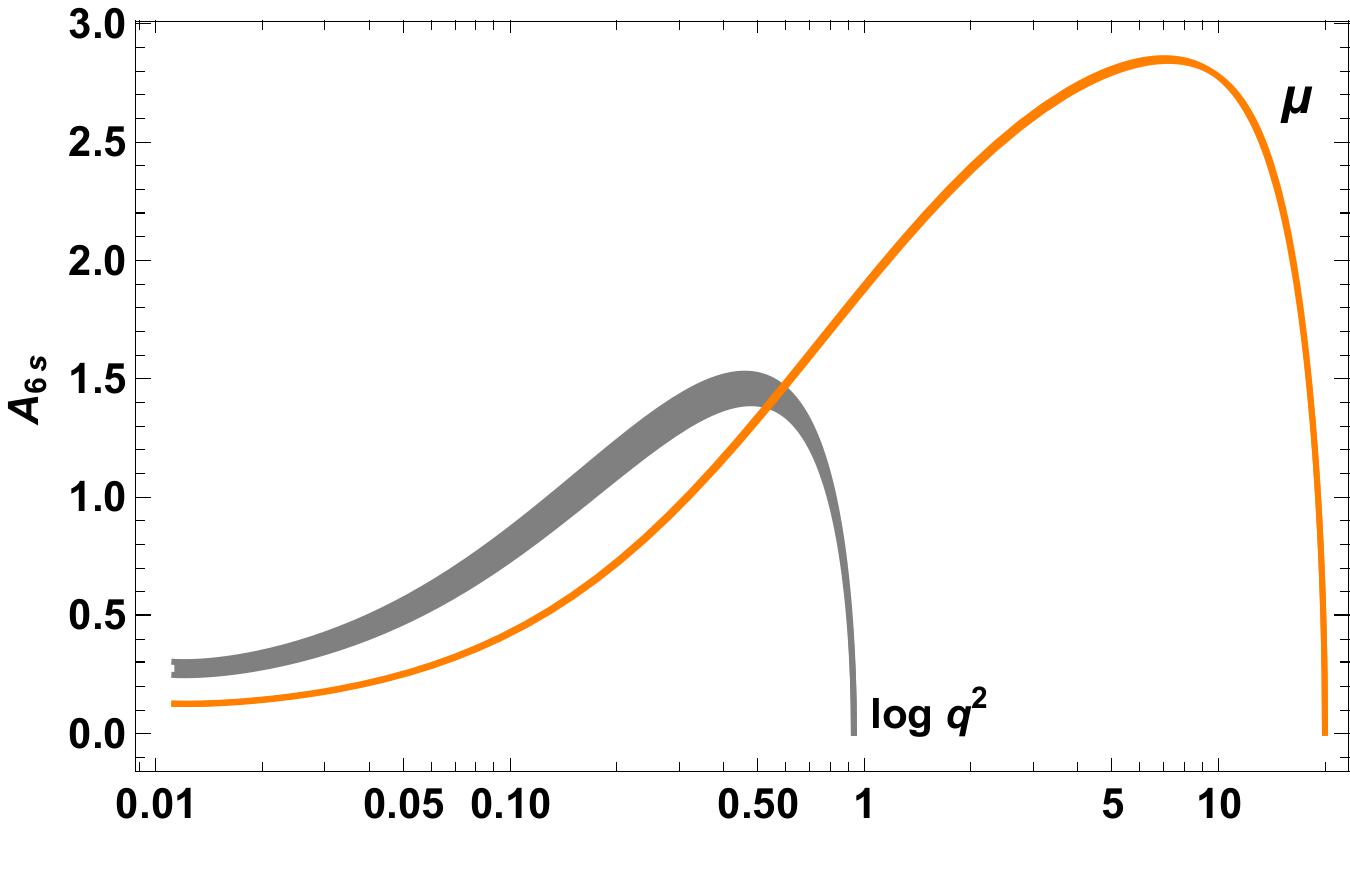}
\caption{SM results for the angular observables $A_3$, $A_4$, $A_5$ and $A_{6s}$ for electron and muon for $B^{*+}_c \rightarrow P(\rightarrow P'\,\mu^+\,\nu_{\mu}) 
\,
\ell^{+}\,{\nu}_\ell$ decays. }
\label{angularOtherDecays}
\end{figure}

Having confirmed that $C_{V_L}$ indeed does not contribute to NP effects in the angular observables, we present 
the SM results for these as a function of $\log q^2$ in Figure~\ref{angularOtherDecays} for the sake of completeness. The color 
coding in Figure~\ref{angularOtherDecays} is consistent with Figures.~\ref{BrVsq2 BsDs} and~\ref{BrVsq2 BcD0} in that we show the 
SM result for $\left[ B_s^0(\rightarrow D_s^{*-}\,\mu^+\,\nu_{\mu}) 
\right]\,
\ell^{+}\,{\nu}_\ell$ in gray and for $\left[ D^0(\rightarrow K^-\,\mu^+\,\nu_{\mu}) 
\right]\,
\ell^{+}\,{\nu}_\ell$ in orange. 

\begin{table}[H]
\caption{Form factors for $B^*_c \to B_s$ and $B_c^* \to D$ transitions with uncertainties \cite{Wang:2024cyi}. The first uncertainty represents statistical errors, while the second corresponds to systematic uncertainties.}
\centering
\resizebox{\textwidth}{!}{
\begin{tabular}{|c|c|c|c|c|c|c|c|c|c|}
\hline
\multicolumn{4}{|c|}{$B^*_c \to B_s$}  & \multicolumn{3}{c|}{$B_c^* \to D$} \\ 
\hline
\textbf{$F(q^2)$} & \textbf{$F(0)$}  & \textbf{$a$}& \textbf{$b$}  & \textbf{$F(0)$}  & \textbf{$a$} & \textbf{$b$} \\
\hline
$V(q^2)$ & $3.61^{+0.09+0.11}_{-0.10-0.14}$   & $5.78^{+0.13+0.10}_{-0.13-0.10}$ &$ 17.20^{+0.02+0.03}_{-0.03-0.03}$  &$0.26^{+0.01+0.01}_{-0.01-0.02}$&$1.58^{+0.30+0.25
}_{-0.27-0.25}$&$1.65^{+0.07+0.07}_{
-0.06-0.07}$ \\
$A_0(q^2)$ & $0.50^{+0.01+0.01}_{-0.02-0.01}$  & $3.70^{+0.13+0.11}_{-0.13-0.11}$ & $7.75^{+0.02+0.02}_{-0.03-0.02}$  &$0.15^{+0.01+0.01}_{-0.01-0.01}$&$1.18^{+0.30+0.27}_{-0.27-0.27}$&$0.82^{+0.06+0.08}_{-0.07-0.06}$\\
$A_1(q^2)$ &$0.52^{+0.00+0.00}_{-0.00-0.01}$   & $4.01^{+0.12+0.12}_{-0.12-0.12}$ & $7.98^{+0.01+0.02}_{-0.02-0.02}$  &$0.16^{+0.01+0.01}_{-0.01-0.01}$&$1.10^{+0.28+0.24}_{-0.25-0.23}$ &$0.70^{+0.06+0.06}_{-0.04-0.05}$\\
$A_2(q^2)$ &$0.27^{+0.17+0.16}_{-0.21-0.16}$   & $-0.90^{+0.11+0.11}_{-0.10-0.11}$& $7.57^{+0.01+0.02}_{-0.01-0.02}$  &$0.13^{+0.01+0.00}_{-0.01-0.01}$& $1.33^{+0.26+0.23}_{-0.24-0.23}$&$1.21^{+0.04+0.06}_{-0.04-0.05}$\\ \hline\hline
\end{tabular}}
\label{ff table 2}
\end{table}

\section{Conclusion}\label{section V}

We have studied the FCCC processes focusing on the 
cascade decay $B^{*0}_{s} \rightarrow 
   D_s^-(\rightarrow \tau^-\,\bar\nu_{\tau})\,
\ell^{+}\,{\nu}_\ell$. We compute this process using weak effective theory in the SM using the 
helicity amplitude formalism and compute the branching ratio, forward-backward asymmetry and 
a host of angular observables. Furthermore, we consider left and right handed chiral vector-like so-called NP effects beyond the SM which are easily incorporated in the weak 
effective Hamiltonian. 

We have shown that left and right handed vector like couplings, $C_{V_L}$ and $C_{V_R}$, both 
contribute to NP effects in the branching ratio but with opposite sign and so can be 
easily distinguished from each other. In both cases the change from the SM is appreciable 
within the $1\sigma$ allowed range of the couplings. 

We show, however, that $C_{V_L}$ does not contribute to NP effects in the 
forward-backward asymmetries and angular observables at the tree level in the weak effective theory in this decay. This finding should be visible in other FCCC processes as well, a fact that 
we confirm that this is a general feature of charged current processes with a vector meson going to a psedoscalar at the tree level in effective weak theory by cross checking with the cascade decays $B^{*+}_c \rightarrow P(\rightarrow P'\,\mu^+\,\nu_{\mu}) 
\,
\ell^{+}\,{\nu}_\ell$ where $P$ is $B_s^0$ ($D^0$) and $P'$ is $D_s^{*-}$ ($K^-$). 
Our overall conclusion is that FCCC is an important channel from the point of view of 
investigating NP effects beyond the SM and may be important as more experimental 
data becomes available.

\providecommand{\href}[2]{#2}\begingroup\raggedright\endgroup

\end{document}